\RequirePackage{fix-cm}

\documentclass[smallextended]{svjour3}       
\smartqed  %
\usepackage{graphicx}
\usepackage[dvipsnames,svgnames]{xcolor}
\definecolor{Mustard}{rgb}{0.21, 0.27, 0.31}
\usepackage{tikz}
\usepackage{pgfplots}
\pgfplotsset{compat=1.7}
\usepackage{subcaption}
\usepackage[mathletters]{ucs}
\usepackage[utf8x]{inputenc}
\usepackage[linesnumbered,ruled,vlined]{algorithm2e}
\usepackage{comment}
\usepackage{booktabs} 
\usepackage[numbers]{natbib}
\usepackage{amsmath}
\usepackage{array, tabularx, ltablex, caption}
\usepackage{enumitem}
\begin{document}

\title{A Privacy-Preserving Model based on Differential Approach for Sensitive Data in Cloud Environment
}

\author{Ashutosh Kumar Singh         \and
         Rishabh Gupta 
}

\author{Ashutosh Kumar Singh \and
Rishabh Gupta}
\authorrunning{Multimedia Tools and Applications}

\institute{Department of Computer Applications, National Institute of Technology, Kurukshetra, Haryana, INDIA\\
\email{ashutosh@nitkkr.ac.in}\\
\email{rishabhgpt66@gmail.com}\\
This article has been accepted in Springer Multimedia Tools and Applications Journal \(\copyright\) 2022 Springer Nature. Personal use of this material is permitted. Permission from Springer must be obtained for all other uses in any current or future media, including reprinting/republishing this material for advertising or promotional purposes, creating new collective works, for resale or redistribution to servers or lists, or reuse of any copyrighted component of this work in other works. This work is freely available for survey and citation.}

\date{}

\maketitle

\begin{abstract}
A large amount of data and applications need to be shared with various parties and stakeholders in the cloud environment for storage, computation, and data utilization. Since a third party operates the cloud platform, owners cannot fully trust this environment. However, it has become a challenge to ensure privacy preservation when sharing data effectively among different parties. This paper proposes a novel model that partitions data into sensitive and non-sensitive parts,  injects the noise into sensitive data, and performs classification tasks using k-anonymization, differential privacy, and machine learning approaches. It allows multiple owners to share their data in the cloud environment for various purposes. The model specifies communication protocol among involved multiple untrusted parties to process owners’ data. The proposed model preserves actual data by providing a robust mechanism. The experiments are performed over Heart Disease, Arrhythmia, Hepatitis, Indian-liver-patient, and Framingham datasets for Support Vector Machine, K-Nearest Neighbor, Random Forest, Naive Bayes, and Artificial Neural Network classifiers to compute the efficiency in terms of accuracy, precision, recall, and F1-score of the proposed model. The achieved results provide high accuracy, precision, recall, and F1-score up to 93.75\%, 94.11\%, 100\%, and 87.99\% and improvement up to 16\%, 29\%, 12\%, and 11\%, respectively, compared to previous works.
\keywords{Machine learning \and Cloud computing \and Differential privacy \and  Laplace distribution \and k-anonymity}
\end{abstract}
\section{Introduction}
Data storage, computation, utilization, analysis, and sharing are the vital required services for any organization to improve performance \cite{R1}. Numerous applications and data are shifting from the local to the cloud due to various benefits such as minimum cost, maximum efficiency, and high scalability \cite{R2}. Sensitive sample data also are shared with the cloud or other parties for distinguish services \cite{R30}. However, users hesitate to share data with the cloud for computation and storage since a third party manages it \cite{R31}, and data may be misused as well as owners lose control of their data \cite{R3} \cite{R4}. The cloud may also provide outsourced data to other entities for different purposes \cite{R5}. Due to these reasons, data protection has become a critical challenge for any organization. Therefore, there is a need for a mechanism that can protect sensitive data. For this, the different kind of techniques, such as cryptography, differential privacy, k-anonymity, etc., are used to preserve the data for privacy reasons before transferring it to the cloud platforms \cite{R6} \cite{R7}. 
\par
To address the aforementioned challenges, we propose a novel Privacy-Preserving Model based on Differential approach (PPMD) for sensitive data in the cloud environment. In the proposed model, owners partition their data into sensitive \& non-sensitive \cite{R21}-\cite{R23}, and different statistical noise is injected into sensitive data according to various applications and owner's queries \cite{R24}-\cite{R26}. Differential privacy protection is considered on the owner's side because they do not want to share actual data. The resulted data is uploaded to the cloud platform, and classification services are provided \cite{R27}. The machine learning algorithms are applied over resulted data for classification. The cloud platform obtains classified data from the classification model and sends it to the data owner rather than other parties. Fig. 1 presents a bird-eye view of the proposed work and highlights our consecutive contributions to preserve data privacy and perform classification tasks in the cloud environment. The summary of the main contributions of PPMD are as follows: 
\begin{itemize}
\item[$\bullet$] PPMD allows various data owners to share outsourced data securely. To protect data against stealing or leakage, noise is injected, and noise-added data is shared.
\item[$\bullet$] PPMD uses the cloud platform for storage, computation, and performing classification tasks over collected data from multiple owners. All entities are considered to be untrusted to protect data with enhanced privacy.
 \item[$\bullet$] PPMD maintains the degree of accuracy because of the data partition into sensitive \& non-sensitive parts and statistical noise addition.
\item[$\bullet$] A series of experiments are conducted using the distinct dataset to validate the practicality of the proposed model. Besides,  the comparisons are performed among the various a) datasets, b) classifiers, and c) distinctly preprocessed data using differential privacy and with the state-of-the-art works to prove the superiority of PPMD.
\end{itemize}
\label{intro}
 	\begin{figure*}[!htbp]
	\hspace{1cm}
	\includegraphics[width=1.0\textwidth]{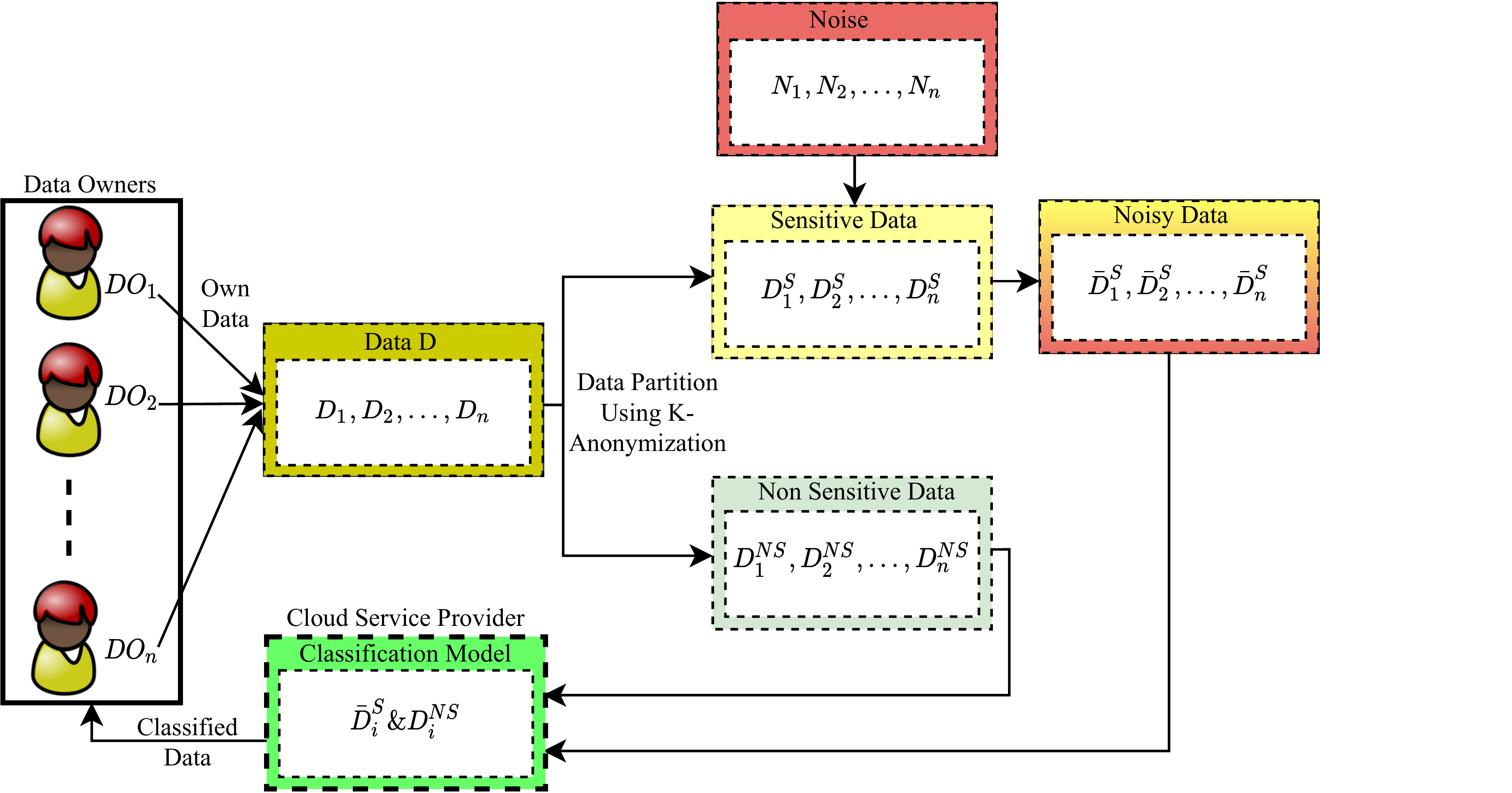}
		\caption{Bird eye view of the proposed work}
		\label{fig:BEW}
\end{figure*}
\textit{Organization:} The rest of this paper proceeds as follows. Section 2 describes the related work. In Section 3, we present the proposed model PPMD, and the process of data partition is shown in Section 4. The data classification steps are defined in Section 5. Section 6 shows the implementation and evaluates the results of the experiment with statistical analysis. Finally, a conclusion and the direction of the future work are described in Section 7. The list of notations with their definitions is shown in Table 1 used in this paper.  
\begin{table}[!htbp]
		\caption{List of Terminologies with their Explanatory Terms}
		\label{TableExp11}
		\begin{center}
			\begin{tabular}{ll|ll}\hline 
                   \hline
				\textit{$DO_{id}$}: & Data Owners & \textit{$CSP$}: & Cloud Service Provider\\
				\textit{$P_{id}$}: & Patients & \textit{$D_{i}$}: & Actual data \\ 
				\textit{$N_{i}$}: & Noise & \textit{$CM$}: & Classification Model \\ 
				\textit{$D_{i}^{S}$}: & Sensitive data & \textit{$CA$}: & Classification Accuracy \\
				\textit{$D_{i}^{NS}$}: & Non-sensitive data & \textit{$P$}: & Precision \\
				\textit{$\bar{D}_{i}^{S}$}: & Noise-added data & \textit{$\hat{D_{i}}$}: & Preprocessed data \\
				\textit{$R$}: & Recall & \textit{$\hat{D}_{t^{'}}$}: & Training data \\
				\textit{$\hat{D}_{t^{''}}$}: & Testing data & \textit{$FS$}: & F1-score \\
				\textit{$A_{i}$}: & Data attribute & \textit{$s^{'}$} & Scaling parameter\\
				\textit{$\mu$}: & Mean & \textit{$\sigma$}: & Standard deviation \\
				\textit{$\epsilon$}: & Privacy budget & \textit{$n^{*}$}: & Count of classes \\
				\textit{$n$}: & Number of data objects & \textit{$L$}: & Label item\\
				\textit{$C$}: & Object category & &\\
		        \hline \hline
			\end{tabular}
		\end{center}
	\end{table}
\section{Related work}
 Yuan and Yu \cite{R8} proposed a secure, efficient, and accurate multiparty Back-Propagation Neural (BPN) network-based scheme that allowed two or more parties, each having an arbitrarily partitioned data set, to conduct the learning collaboratively. But, this scheme focused on facilitating data processing without considering the algorithm efficiency. Yonetani et al. \cite{R9} proposed a doubly permuted homomorphic encryption (DPHE) based privacy-preserving framework. It enabled multiparty protected scalar products with a reduction of the high computational cost. However, DPHE supports either addition or multiplications at a particular time. A system that utilizes additively homomorphic encryption to protect the gradients against the curious server is presented in \cite{R10}. This system achieved identical accuracy to a corresponding deep learning system, i.e., asynchronous stochastic gradient descent (ASGD) trained over the common dataset of all participants. All classifiers from multiple parties are trained over the single-source domain in this scheme, but the trade-off is a lower accuracy rate. To provide the privacy-preserving classification service for users, Li et al. \cite{R11} proposed a scheme for a classifier owner to delegate a remote server. But during the launch of a classification query, user interactions were often involved in this scheme. Li et al. \cite{R12} proposed a data protection scheme that enables a trainer to train a Naive Bayes classifier over the dataset provided jointly by the different data owners. To preserve the privacy of the data, $\epsilon$-differential privacy is utilized. But, adversaries still have the ability to forge and manipulate the data in this scheme. Ma et al. \cite{R13} proposed a privacy-preserving deep learning model, namely PDLM, to train the model over the encrypted data under multiple keys. A privacy-preserving calculation toolkit was adopted to train the model based on stochastic gradient descent (SGD) in a privacy-preserving manner. The model reduced the storage overhead, but the classification accuracy is less, and the computation cost is high. Li et al. \cite{R14} proposed a privacy-preserving machine learning with multiple data provider (PMLM) scheme with improved computational efficiency and data analysis accuracy. The authors used public-key encryption with a double decryption algorithm (DD-PKE) and $\epsilon$-differential privacy. But the scheme suffered from less accuracy as well as less data sharing. To protect the confidentiality of sensitive data without leakage, a privacy-conserving outsourced classification in cloud computing (POCC) framework was introduced \cite{R15} under various public keys using a fully homomorphic encryption proxy technique. However, data owners and storage servers are deemed to be in the same trustworthy domain that no longer exists in the cloud environment. Gao et al. \cite{R16} proposed a scheme to avoid information leakage under the substitution-then-comparison (STC) attack. A privacy-preserving classification mechanism was designed by adopting a double-blinding technique for Naive Bayes, and both the communication and computation overhead was reduced. But this scheme is unable to obtain the discovery of truth that protects privacy. To apply the deep neural network algorithms over the encrypted data, Hesamifard et al. \cite{R17} revert a neural networking-based framework named CryptoDL while considering the existing limitations of homomorphic encryption schemes. Although the approach works well to secure private data, but the data is protected using a key that is not feasible. Phong and Phuong \cite{R18} constructed a privacy-preserving system, namely the server-aided network topology (SNT) system, and the fully-connected network topology (FNT) system, depending on the connection with SNT and FNT server. In these systems, multiple machine-learning trainers can use the SGD or its variants over the combined dataset without sharing the local dataset of each trainer. The constructed systems used the weight parameters rather than the gradient parameters and achieved an accuracy similar to SGD. 
 Wei et al. \cite{R29} proposed a framework, namely, noising before model aggregation federated learning (NbAFL), which prevents the information leakage effectively. The client's data is protected by using a differential privacy mechanism. However, this framework requires a large amount of noise to add and sacrificing mode utility.
 Gupta et al. \cite{R19} proposed a machine learning and probabilistic analysis-based model, namely MLPAM. It supports multiple participants to share their data safely for different purposes by using encryption, machine learning, and probabilistic approaches. The proposed model provided a mechanism that reduced the risk associated with the leakage for prevention coupled with detection. The experimental results showed that the proposed model ensured high accuracy and precision. A summary of the literature review is depicted in Table 2.
{\setlength\defaultaddspace{0.5ex}

   \captionof{table}{Tabular sketch of the literature review}
\small
\begin{tabularx}{\linewidth}{|p{1.8cm}|p{3.5cm}|p{3cm}|p{1.7cm}|}

\hline
        Model/Scheme /Framework & Workflow \& Implementation & Outcomes & Drawback \\
        \hline  \hline
        A secured scheme for processing ciphered text \cite{R8} 
        & \vspace{-0.3cm} 
        \begin{itemize}[topsep=-0.3cm,leftmargin=0.2cm,label=\textbullet]
            \item The arbitrarily partitioned data is encrypted using a doubly homomorphic encryption scheme 
            \item Experiments are per-formed on the Amazon EC2 cloud 
        \end{itemize} & 
       \vspace{-0.3cm} 
        \begin{itemize}[topsep=-0.3cm,leftmargin=0.2cm,label=\textbullet]
            \item Security analysis proves that this scheme is secure, scalable, and efficient
            \item Less error rates
        \end{itemize} & 
        High computation and communication complexity \\
        \hline
        A privacy-preserving framework for visual learning \cite{R9} 
        & 
        \vspace{-0.3cm} 
        \begin{itemize}[topsep=-0.3cm,leftmargin=0.2cm,label=\textbullet]
            \item A homomorphic cryptosystem is used to update high-dimensional classifiers 
            \item The experiments are performed on the CelebA and Life-logging datasets 
        \end{itemize} & 
      \vspace{-0.3cm} 
        \begin{itemize}[topsep=-0.3cm,leftmargin=0.2cm,label=\textbullet]
            \item Achieve 84\% accuracy by performing facial recognition tasks
            \item Minimize the computational cost of homomorphic encryption
        \end{itemize} & 
        Does not assist the multiple operations (addition or multiplication) \\
        \hline
       A secure deep learning system for parameters protection \cite{R10}
        & 
        \vspace{-0.3cm} 
        \begin{itemize}[topsep=-0.3cm,leftmargin=0.2cm,label=\textbullet]
            \item The homomorphic encryption and asynchronous stochastic gradient are adopted to encrypt trained parameters
            \item The Adam optimizer is used for training with input learning rate $10^{−4}$
        \end{itemize} & 
        \vspace{-0.3cm} 
        \begin{itemize}[topsep=-0.3cm,leftmargin=0.2cm,label=\textbullet]
            \item More effective in protecting sensitive information from the curious server
            \item Achieve the same accuracy as that of the centralized DL algorithm
        \end{itemize} & 
        The local data can still be surreptitiously extracted from two adjacent versions of parameters \\
        \hline
   A secure outsourcing scheme for the classification service \cite{R11}
        & 
       \vspace{-0.3cm} 
        \begin{itemize}[topsep=-0.3cm,leftmargin=0.2cm,label=\textbullet]
            \item The additive homomorphic encryption technique is used to protect the data
            \item The experiments are conducted on the LAN server
        \end{itemize} & 
       \vspace{-0.3cm} 
        \begin{itemize}[topsep=-0.3cm,leftmargin=0.2cm,label=\textbullet]
            \item More practical and less communication cost
        \end{itemize} & 
        Only support a single-party setting \\
        \hline
  A privacy-preserving scheme for learning algorithms \cite{R12}
        & 
       \vspace{-0.3cm} 
        \begin{itemize}[topsep=-0.3cm,leftmargin=0.2cm,label=\textbullet]
            \item The differential privacy mechanism was used to protect the data
            \item The LAN server was used to conduct the experiments
        \end{itemize} & 
      \vspace{-0.3cm} 
        \begin{itemize}[topsep=-0.3cm,leftmargin=0.2cm,label=\textbullet]
            \item Achieve data privacy will not break while the training data
            \item Less computational time requires for training
        \end{itemize} & 
        Paillier cryptosystem can only work with integers \\
        \hline
  A privacy preserving deep learning model to train over the encrypted data \cite{R13}
        & 
       \vspace{-0.3cm} 
        \begin{itemize}[topsep=-0.3cm,leftmargin=0.2cm,label=\textbullet]
            \item The model is trained based on stochastic gradient descent, the feed-forward, and a back-propagation procedure
            \item The experiments were conducted over MNIST, CIFAR-10 datasets
        \end{itemize} & 
     \vspace{-0.3cm} 
        \begin{itemize}[topsep=-0.3cm,leftmargin=0.2cm,label=\textbullet]
            \item Minimize the storage overhead
            \item Errors are calculated to evaluate the performance of the model
        \end{itemize} & 
         Low efficiency utilizing the distributed two trapdoors public-key cryptosystem \\
        \hline
  A privacy-preserving machine learning scheme for data protection \cite{R14}
        & 
    \vspace{-0.3cm} 
        \begin{itemize}[topsep=-0.3cm,leftmargin=0.2cm,label=\textbullet]
            \item A double encryption algorithm and differential privacy mechanism was used to preserve data privacy
            \item MAGMA programming is used to perform the cryptosystem
        \end{itemize} & 
      \vspace{-0.3cm} 
        \begin{itemize}[topsep=-0.3cm,leftmargin=0.2cm,label=\textbullet]
            \item Enhance the computational efficiency and data analysis accuracy
            \item The security analysis proves that the model is more secure
        \end{itemize} & 
       High computational cost due to the dependence on integer factorization \\
        \hline
A privacy-preserving outsourced classification framework for confidentiality of sensitive data \cite{R15}
        & 
    \vspace{-0.3cm} 
        \begin{itemize}[topsep=-0.3cm,leftmargin=0.2cm,label=\textbullet]
            \item A fully homomorphic encryption proxy technique is utilized to encrypt data
            \item The naive bayes classifier is performed over multiple datasets for experiment work
        \end{itemize} & 
      \vspace{-0.3cm} 
        \begin{itemize}[topsep=-0.3cm,leftmargin=0.2cm,label=\textbullet]
            \item Reduce power consumption by cloud clusters
            \item Less computational overhead
        \end{itemize} & 
       The data is encrypted with a single key, and it is not suitable for multi-user systems \\
        \hline
A privacy-preserving Naive Bayes classifier scheme to prevent information leakage \cite{R16}
        & 
    \vspace{-0.3cm} 
        \begin{itemize}[topsep=-0.3cm,leftmargin=0.2cm,label=\textbullet]
            \item A double-blinding technique is used to avoid the attacks
            \item The GNU Multi-Precision library and the OpenSSL library are used for programming
        \end{itemize} & 
      \vspace{-0.3cm} 
        \begin{itemize}[topsep=-0.3cm,leftmargin=0.2cm,label=\textbullet]
            \item Reduce the computation cost because of not using fully homomorphic encryptions
            \item More efficient due to offline phase of server
        \end{itemize} & 
       Not suitable for the multi-label dataset \\
        \hline
A CryptoDL framework for applying deep neural network algorithms over encrypted data \cite{R17}
        & 
    \vspace{-0.3cm} 
        \begin{itemize}[topsep=-0.3cm,leftmargin=0.2cm,label=\textbullet]
            \item The polynomial approximation is used for continuous activation functions
            \item The experiments are conducted on MNIST and CIFAR-10 datasets
        \end{itemize} & 
      \vspace{-0.3cm} 
        \begin{itemize}[topsep=-0.3cm,leftmargin=0.2cm,label=\textbullet]
            \item Perform proper privacy-preserving training and classification tasks
            \item The security model proves that the server cannot access the input data
        \end{itemize} & 
       High computation cost and less accuracy \\
        \hline
A privacy-preserving deep learning model for input privacy \cite{R18}
        & 
    \vspace{-0.3cm} 
        \begin{itemize}[topsep=-0.3cm,leftmargin=0.2cm,label=\textbullet]
            \item The input data is protected through symmetric encryption
            \item A multilayer perceptron and a convolutional neural network are used for experiments 
        \end{itemize} & 
      \vspace{-0.3cm} 
        \begin{itemize}[topsep=-0.3cm,leftmargin=0.2cm,label=\textbullet]
            \item Achieve the same learning accuracy as SGD
            \item The security analysis proves that no training information will be revealed using the weight parameters 
        \end{itemize} & 
       To update weight causes low efficiency and no consideration for output privacy \\
        \hline
A privacy-preserving framework for data protection \cite{R29}
        & 
    \vspace{-0.3cm} 
        \begin{itemize}[topsep=-0.3cm,leftmargin=0.2cm,label=\textbullet]
            \item To preserve privacy of data, the Gaussian noise is added to it
            \item The experiments are performed on the MNIST dataset using multi-layer perception
        \end{itemize} & 
      \vspace{-0.3cm} 
        \begin{itemize}[topsep=-0.3cm,leftmargin=0.2cm,label=\textbullet]
            \item Maintain a privacy level with a low-performance loss
            \item  Achieve high level privacy-preserving and secure capabilities
        \end{itemize} & 
        Noise inevitably reduces the accuracy \\
        \hline
A machine learning and probabilistic analysis-based model for secure sharing data \cite{R19}
        & 
    \vspace{-0.3cm} 
        \begin{itemize}[topsep=-0.3cm,leftmargin=0.2cm,label=\textbullet]
            \item The differential privacy, encryption, machine learning, and probabilistic approaches are used to encrypt, noise addition, and share the data of multiple participants
            \item The SVM, Random Forest, KNN, and Naive Bayes classifiers train the model on Glass, Iris, Wine, and Balance Scale datasets
        \end{itemize} & 
      \vspace{-0.3cm} 
        \begin{itemize}[topsep=-0.3cm,leftmargin=0.2cm,label=\textbullet]
            \item Minimize the risk affiliated with the leakage for prevention and detection
            \item Achieve high accuracy and precision up to 97\% and 100\%
        \end{itemize} & 
       Does not provide efficient data sharing and management in multiple environments \\
        \hline
\end{tabularx}
\label{table:ressuffixes11}
}
\par
The major limitations of the existing works are that the models injected the noise into entire data and/or protected it using various encryption approaches followed by machine learning-based classification, which reduced accuracy and/or increased computation cost. The earlier models considered a single owner and/or a single untrusted entity. Unlike the previous works, PPMD partitions the data using k-anonymization, injecting the noise into the sensitive part of data to make it private and applying various state-of-art classifiers. It also permits multiple owners to share outsourced data securely while treating all participating entities involved as untrustworthy.
 \section{Proposed model} The proposed model architecture, named PPMD (Fig. 2), comprises the involved entities and their communication with essential flows.
 	\begin{figure*}[!htbp]
	\begin{center}
	\includegraphics[width=0.95\textwidth]{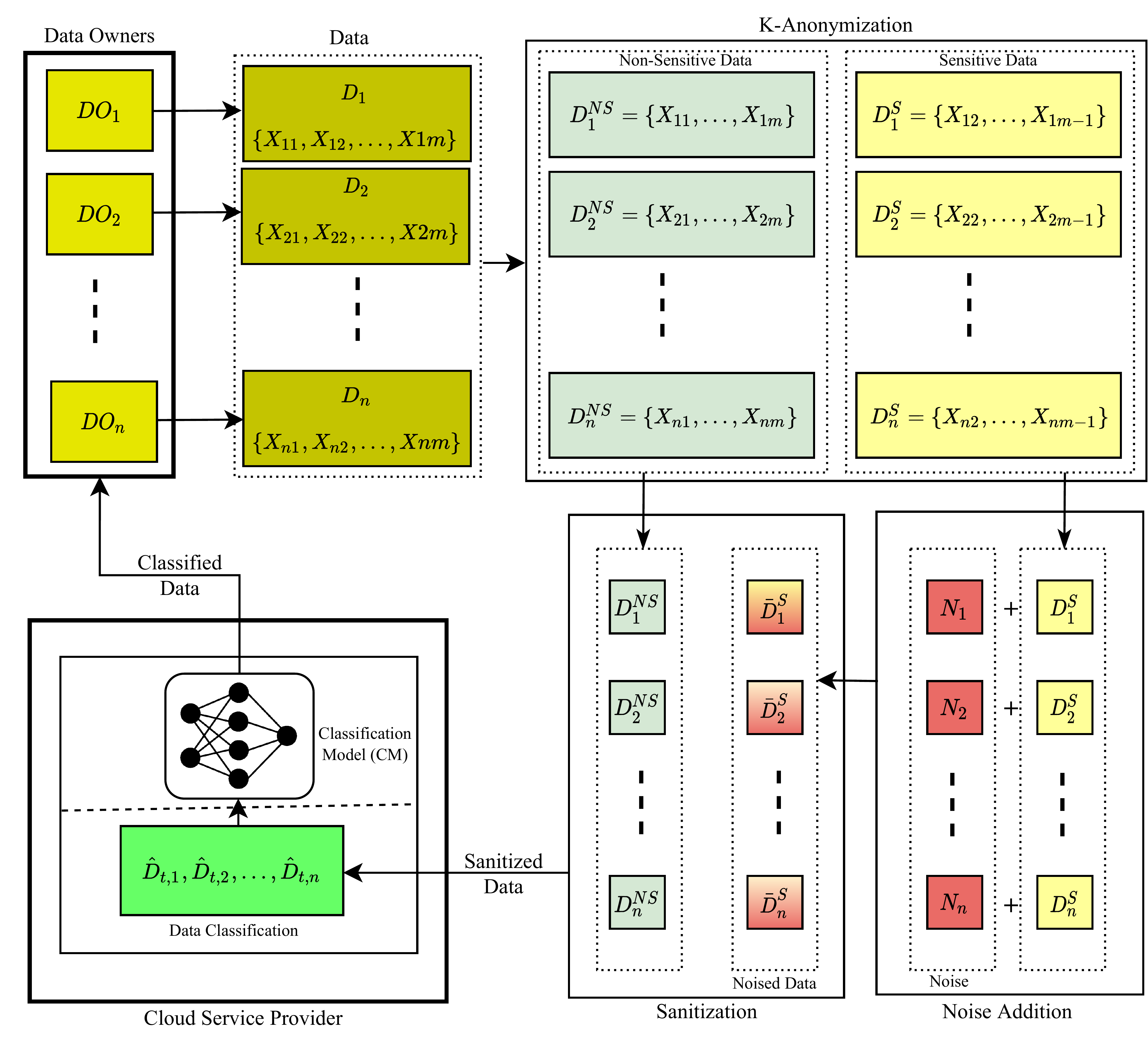}
		\caption{Proposed PPMD architecture}
		\label{fig:PPMD}
	\end{center}
\end{figure*}
 This architecture contains two entities Data Owners (\textit{$DO_{id}$}) and Cloud Service Provider (\textit{$CSP$}) which are described as follow:
 \begin{enumerate}
 \item[1)]\textit{$DO_{id}$}: An entity that generates data and requests to \textit{$CSP$} for services. \textit{$DO_{id}$} sends sensitive/non-sensitive data, but noise is added into sensitive data before transferring it to \textit{$CSP$}. \textit{$DO_{id}$} applies $\epsilon$-differential privacy to protect sensitive data. Since it is assumed that \textit{$DO_{id}$} can't leak its data but may leak other owners' data. Therefore, \textit{$DO_{id}$} is considered an untrusted entity.
 \item[2)] \textit{$CSP$}: An entity that gathers all data from \textit{$DO_{id}$} and provides facilities of storage and computation. It offers classification services to \textit{$DO_{id}$} through the classification model (\textit{CM}). It trains \textit{CM} using machine learning algorithms over collected data and accesses classified data from \textit{CM}. These obtained results are shared among  \textit{$DO_{id}$}. In PPMD, \textit{$CSP$} is treated as a semi-trusted entity, as it strictly follows the protocol but is curious to learn the information.
 \end{enumerate}
 Let the data owners \textit{$DO_{id}$} = \{$DO_{1}$, $DO_{2}$, $\dots$, $DO_{n}$\} has data \textit{$D$} = \{$D_{1}$, $D_{2}$, $\dots$, $D_{n}$\}, where the data object $D_{i} \in \textit{D}$ is independent and can be of any type and size. \textit{$DO_{id}$} needs to share \textit{$D$} with the semi-trusted party like \textit{$CSP$} for storage, computation, and performance enhancement. But, \textit{$DO_{id}$} can't share $D$ because it contains sensitive data also. 
 Therefore, before sharing, \textit{$DO_{id}$} partitions his data into sensitive data $D^{S}$ and non-sensitive data $D^{NS}$ using the k-anonymization mechanism. In order to make $D^{S}$ private, \{$DO_{1}$, $DO_{2}$, $\dots$, $DO_{n}$\} procure noisy data $\bar{D}^S$ = \{$\bar{D}_{1}^S, \bar{D}_{2}^S, $\dots$, \bar{D}_{n}^S$\} by adding noise $N$ = \{$N_{1}$, $N_{2}$, $\dots$, $N_{n}$\} into $D^{S}$ = \{$D_{1}^{S}$, $D_{2}^{S}$, $\dots$, $D_{n}^{S}$\} using the $\epsilon$-differential privacy. \textit{$DO_{id}$} has noise-added data $\bar{D}^S$ \& non-sensitive data $D^{NS}$ that are combined, and sanitized data $\hat{D}= \{\hat{D}_{t,1},\hat{D}_{t,2},\dots,\hat{D}_{t,n}\}$ is obtained. \textit{$DO_{id}$} sends $\hat{D}$ to \textit{$CSP$} that performs the classification tasks over it to make a fit \textit{CM}. Any query can be made by $DO_{1}$, $DO_{2}$, $\dots$, $DO_{n}$ to \textit{$CSP$}. The results of these queries are received by \textit{$CSP$} from \textit{CM}, and sent to the corresponding entity $DO_{1}$, $DO_{2}$, $\dots$, $DO_{n}$. Algorithm 1 shows the operational summary of the proposed model. Initially, data $D_{i}$ is divided  into $D_{i}^{S}$ and $D_{i}^{NS}$. Afterward, noise vector $N_{i}$ is generated and performed the addition operation on $D_{i}^{S}$ and $N_{i}$. The classification operation is carried out over noisy \& non-sensitive data using the machine learning algorithms, and unknown class labels are obtained from \textit{CM}. The accuracy, precision, recall, and F1-score are calculated using these class labels.
\begin{algorithm}
\KwIn{Actual data $D$, Sensitive data $D^{S}$, Non-sensitive data $D^{NS}$, Noise vector $N$, input vector $\hat{D}$}
\KwOut{$CA$, $P$, $R$, and $FS$}
Initialize data $D$ := \{$D_{1}$, $D_{2}$, $\dots$, $D_{n}$\}, $\bar{D}^{S}$ := \{$\bar{D}_{1}^{S}$, $\bar{D}_{2}^{S}$, $\dots$, $\bar{D}_{n}^{S}$\}, $N$ := \{$N_{1}$, $N_{2}$, $\dots$, $N_{n}$\}  \\
\textbf{for} $i$ = $1$, $2$, $\dots$, $n$ \textbf{do}\\
\hspace{.3cm}\textbf{Data\_Partition} ($D_{i}$) \\
\hspace{.3cm} $N_{i}$ = Lap(0,1) \\
\hspace{.3cm}$\bar{D}_{i}^{S}$ = $D_{i}^{S}$ + $N_{i}$ \\
\hspace{.3cm}$\hat{D}_{i}$ = ($\bar{D}_{i}^{S}$, $D_{i}^{NS}$) \\
\hspace{.3cm}\textbf{Data\_Classification} ($\hat{D}_{i}$) \\
\textbf{end for} \\
$CA$ = (\#$Correctly$ $classified$ $sample$ /  \#$test$ $sample$) * 100 \\
$P$ = $(TP) / (TP + FP)$\\
$R$ = $(TP) / (TP + FN)$\\
$FS$ = $2 * (P * R) / (P + R)$
\caption{PPMD model operational summary}
\label{algo:bb}
\end{algorithm}
\section{Data partition \& noise addition}
In PPMD, \textit{$DO_{id}$} = \{$DO_{1}$, $DO_{2}$, $\dots$, $DO_{n}$\} has data \textit{$D$} = \{$D_{1}$, $D_{2}$, $\dots$, $D_{n}$\} in the form of relation,
$D_{1}$, $D_{2}$, $\dots$, $D_{n}$ has attributes $\{A_{1}$, $A_{2}$, $\dots$, $A_{t_{1}}\}$, $\{A_{1}$, $A_{2}$, $\dots$, $A_{t_{2}}\}$, $\dots$, $\{A_{1}$, $A_{2}$, $\dots$, $ A_{t_{n}}\}$. These relations \{$D_{1}$, $D_{2}$, $\dots$, $D_{n}$\} are partitioned into two relations sensitive and non-sensitive based on row-level data sensitivity of using k-anonymization approach by applying Eq. (1) and (2). The Algorithm 2 presents the  partition of data $D_{1}$ into $D_{1}^{S}$ = \{$A_{1}$, $A_{2}$, $\dots$, $A_{t_{1}-p}$\} and $D_{1}^{NS}$ = \{$A_{t_{1}-p+1}$, $A_{t_{1}-p+2}$, $\dots$, $A_{t_{1}}$\}, $D_{2}$ is partitioned into $D_{2}^{S}$ = \{$A_{1}$, $A_{2}$, $\dots$, 
$A_{t_{2}-q}$\} and $D_{2}^{NS}$ = \{$A_{t_{2}-q+1}$, $A_{t_{2}-q+2}$, $\dots$, $A_{t_{2}}$\}, $\dots$, $D_{n}$ is partitioned into $D_{n}^{S}$ = \{$A_{1}$, $A_{2}$, $\dots$, $A_{t_{n}-r}$\} and $D_{n}^{NS}$ = \{$A_{t_{n}-r+1}$, $A_{t_{n}-r+2}$, $\dots$, $A_{t_{n}}$\} respectively, where $p,q,r,s \in Z$.

 \begin{equation}
   D_{i}^{S} = \prod_{(A_{1}, A_{2}, \dots, A_{t_{i}-s})}\left(D_{i}\right)
  \end{equation}

 \begin{equation}
  D_{i}^{NS} = \prod_{(A_{t_{i}-s+1}, A_{t_{i}-s+2}, \dots, A_{t_{i}})}\left(D_{i}\right)
  \end{equation}

\begin{algorithm}

\KwIn{Actual data $D$ with $\ddot{n}$ records, the value of $\ddot{k}$ for k-anonymity}
\KwOut{ Sensitive data $D^{S}$, Non-sensitive data $D^{NS}$}
Initialize data $D$ := \{$D_{1}$, $D_{2}$, $\dots$, $D_{n}$\}, $D^{S}$ := \{$D_{1}^{S}$, $D_{2}^{S}$, $\dots$, $D_{n}^{S}$\}, $D^{NS}$ := \{$D_{1}^{NS}$, $D_{2}^{NS}$, $\dots$, $D_{n}^{NS}$\}  \\
Set $\ddot{p}$ = $\left\lfloor\frac{\ddot{n}}{\ddot{k}}\right\rfloor$  \\ 

\textbf{for} $i$ = $1$, $2$, $\dots$, $n$ \textbf{do}\\
\hspace{.3cm} \textbf{for} $\ddot{e}$ = $1$, $\dots$, $\ddot{p}$ \textbf{do}\\
\hspace{.58cm} Randomly select distinct records $r_{\ddot{e}}$ $\in$ $D_{i}$  \\
\hspace{.7cm}\textbf{while} ($D_{i}$ $\neq$ $\phi$) \textbf{do}\\
\hspace{1.15cm}Add the records $r_{\ddot{e}}$ to $D_{i}^{S}$ \\
\hspace{1.1cm}$D_{i}^{NS}$ = $D_{i}$ $\setminus$ \{$r_{\ddot{e}}$\} \\
\hspace{.7cm}\textbf{end while} \\
\hspace{.3cm}\textbf{end for} \\
\hspace{.3cm}return $D_{i}^{S}$, $D_{i}^{NS}$  \\
\textbf{end for} \\
\caption{Data Partition}
\label{algo:bc}
\end{algorithm}
\setcounter{table}{2}
\begin{table}[!ht]
    \centering
\begin{minipage}[t]{0.48\linewidth}\centering
\caption{Patients Report}
\begin{tabular}{ c c c c c c}
\toprule
Age & Gender & TB & DB & ALG & Disease   \\
\midrule
65 & Male & 0.7 & 0.1 & 3.3 & 1\\ 
62 & Female & 10.9 & 5.5 & 0.2 & 0 \\
68 & Female & 7.3 & 4.1 & 3.3 & 1 \\
58 & Male & 3.9 & 2.0 & 0.4 & 1 \\
72 & Female & 3.2 & 3.7 & 3.4 & 0 \\
46 & Male & 6.4 & 1.0 & 2.2 & 1 \\
 \vdots &\vdots &\vdots &\vdots &\vdots &\vdots \\
    &      &     &  & & \\
34 & Female & 4.6 & 2.4 & 3.1 & 1 \\
\bottomrule
\end{tabular}
\end{minipage}\hfill %
\\
\vspace{.5cm}

\begin{minipage}[t]{0.38\linewidth}
\centering
\caption{\centering Patients Report1 (Sensitive)}
\label{tab:The parameters 212}
\begin{tabular}{ c c }
\toprule
Age & Gender   \\
\midrule
65 & Male\\ 
62 & Female  \\
68 & Female  \\
58 & Male  \\
72 & Female  \\
46 & Male  \\
 \vdots &\vdots  \\
    &       \\
34 & Female  \\
\bottomrule
\end{tabular}
\end{minipage}
\begin{minipage}[t]{0.48\linewidth}
\centering
\caption{\centering Patients Report2 (Non-sensitive)}
\label{tab:The parameters 246}
\begin{tabular}{ c c c c }
\toprule
TB & DB & ALG & Disease   \\
\midrule
 0.7 & 0.1 & 3.3 & 1\\ 
 10.9 & 5.5 & 0.2 & 0 \\
 7.3 & 4.1 & 3.3 & 1 \\
 3.9 & 2.0 & 0.4 & 1 \\
3.2 & 3.7 & 3.4 & 0 \\
 6.4 & 1.0 & 2.2 & 1 \\
\vdots &\vdots &\vdots &\vdots \\
    &  & & \\
 4.6 & 2.4 & 3.1 & 1 \\
\bottomrule
\end{tabular}
\end{minipage}
\end{table}
For instance, let the proposed PPMD model consist of fifty patients \textit{$P_{id}$} = \{$P_{1}$, $P_{2}$, $\dots$, $P_{50}$\} having data $D$ = \{$D_{1}$, $D_{2}$, $\dots$, $D_{50}$\} in the vector form \{$x_{\ddot{i}}$,$y_{\ddot{i}}$\}, shown in Table 3, which contains both sensitive and non-sensitive data. Therefore, we need to separate the Patients Report into two relations $a)$ Patients Report1 with attributes Age \& Gender, and $b)$ Patient Report2 with attributes TB, DB, ALG, and Disease, correspondingly shown in Tables 4 and 5.  In this way, diseases can't be recognized without knowing Age and Gender. 
\par
 To preserve the privacy of sensitive data, noise is inserted into the tuples of Patients Report1 using differential privacy before transferring to the cloud platform. But, Patients Report2 is outsourced to the same cloud platform without performing any operation. \textit{$DO_{id}$} generates a noise vector $N$ = \{$N_{1}$, $N_{2}$, $\dots$, $N_{n}$\} using the probability density function, and distribution function using Eq. (3). 
\begin{equation}
    N = \frac{1}{2s^{'}} \cdot (exp(\frac{-|rn|}{s^{'}}))
\end{equation}
where $rn$ is input, $s^{'}$ is the scale parameter, and $N$ is a noise vector, which is drawn from the Laplacian distribution with scale $s^{'}$. The generated noise vector $N$ is added in the corresponding data $D^{S}$ = \{$D_{1}^{S}$, $D_{2}^{S}$, $\dots$, $D_{n}^{S}$\} as $\bar{D}_{i}^{S}$ = $D_{i}^{S}$ + $N_{i}$ where $i$ $\in$ [$1$, $n$]. After adding noise, $D^{S}$ data becomes noise-added data $\bar{D}^{S}$ = \{$\bar{D}_{1}^{S}, \bar{D}_{2}^{S}, $\dots$, \bar{D}_{n}^{S}$\} i.e. Patients Report3, shown in Table 6. $\bar{D}^{S}$ and $D^{NS}$ data are combined using the sanitization process, and sanitized data $\hat{D} = \{\hat{D}_{t,1},\hat{D}_{t,2},\dots,\hat{D}_{t,n}\}$ is transferred to \textit{$CSP$}.
 \begin{table}[!ht]
    \centering
\begin{minipage}[t]{0.38\linewidth}
\centering
\caption{\centering Patients Report3 (Noise-added Data)}
\label{tab:The parameters 278}
\begin{tabular}{ c c }
\toprule
Age & Gender   \\
\midrule
65.23 & 1.65\\ 
62.35 & 0.79  \\
68.64 & 0.69  \\
58.75 & 1.49  \\
72.96 & 0.67  \\
46.74 & 1.69  \\
 \vdots &\vdots  \\
    &       \\
34.56 & 0.96  \\
\bottomrule
\end{tabular}
\end{minipage}
\end{table}
\section{Data classification}
\textit{$CSP$} obtains $\hat{D} = \{\hat{D}_{t,1},\hat{D}_{t,2},\dots,\hat{D}_{t,n}\}$ from \textit{$DO_{id}$} and prepossess it by using the normalization function given in Eq. (4), where $T_{s}$ is training sample, $\mu$, and $\sigma$ are the mean and the standard deviation of the training sample, respectively. 
\begin{equation}
    \hat{D} = \frac{(T_s - \mu)}{\sigma}
\end{equation}
It is assumed that data $\hat{D}$= \{$\hat{D}_{t,1}$, $\hat{D}_{t,2}$, $\dots$, $\hat{D}_{t,n}$\} belong to $n^{*}$ $\leq$ $n$ classes $C = \{C_{1}, C_{2}, \dots, C_{n^{*}}\}$ where $n^{*}$ is count of classes, $\bigcup_{i=1}^{n^{*}}$ $C_{i} = D$ and $C_{i} \bigcap C_{j}$ = $\Phi, \forall_{i,j} = 1, 2, \dots, n^{*} \wedge  i \neq j$. 
The steps for the classification of data are illustrated in Fig. 3. In which, data $\{\hat{D}_{t,1},\hat{D}_{t,2},\dots,\hat{D}_{t,n}\}$ is divided into training data $\hat{D}_{t^{'}}$ = \{$\hat{D}_{t,1}$, $\hat{D}_{t,2}$, $\dots$, $\hat{D}_{t,n-k}$\}, and testing data $\hat{D}_{t^{''}}$ = \{$\hat{D}_{t,n-k+1}$, $\hat{D}_{t,n-k+2}$, $\dots$, $\hat{D}_{t,n}$\}. The training data \{$\hat{D}_{t,1},\hat{D}_{t,2},\dots,\hat{D}_{t,n-k}$\} is used to train \textit{$CM$}, whereas accuracy of \textit{$CM$} can be measured by testing data \{$\hat{D}_{t,n-k+1}$, $\hat{D}_{t,n-k+2}$, $\dots$, $\hat{D}_{t,n}$\}. During the testing process, the data object \{$\hat{D}_{t,n-k+1}$, $\hat{D}_{t,n-k+2}$, $\dots$, $\hat{D}_{t,n}$\} is given to \textit{$CM$}, which identifies their classes. \textit{$CM$} analyzes \{$\hat{D}_{t,n-k+1}$, $\hat{D}_{t,n-k+2}$, $\dots$, $\hat{D}_{t,n}$\} and produces a Label vector $L = \{L_{t,n-k+1},L_{t,n-k+2},\dots,L_{t,n}\}$ as an output, whereas $L_{i^{'}} \in L$ specifies $C_{i} \in C$ to which $\hat{D}_{t^{''},i^{'}} \in \hat{D}_{t^{''}}$ pertains. The classification accuracy (\textit{$CA$}) is calculated using Eq. (5), whereas $CI$ indicates the number of items correctly classified and $TI$ indicates the total number of test items. 
\begin{equation}
    CA = \frac{CI}{TI}
\end{equation}
The precision (\textit{$P$}), and recall (\textit{$R$}) are calculated using Eq. (6) and (7) respectively, whereas $TR$ indicates the total number of items returned by the classifier and $RI$ indicates the total number of relevant items. The F1-score (\textit{$FS$}) is measured using Eq. (8).
\begin{equation}
    P = \frac{CI}{TR}
\end{equation}
\begin{equation}
    R = \frac{CI}{RI}
\end{equation}
\begin{equation}
    FS = \frac{2PR}{P+R}
\end{equation}

 	\begin{figure*}[!htbp]
	\begin{center}
	\includegraphics[width=0.95\textwidth]{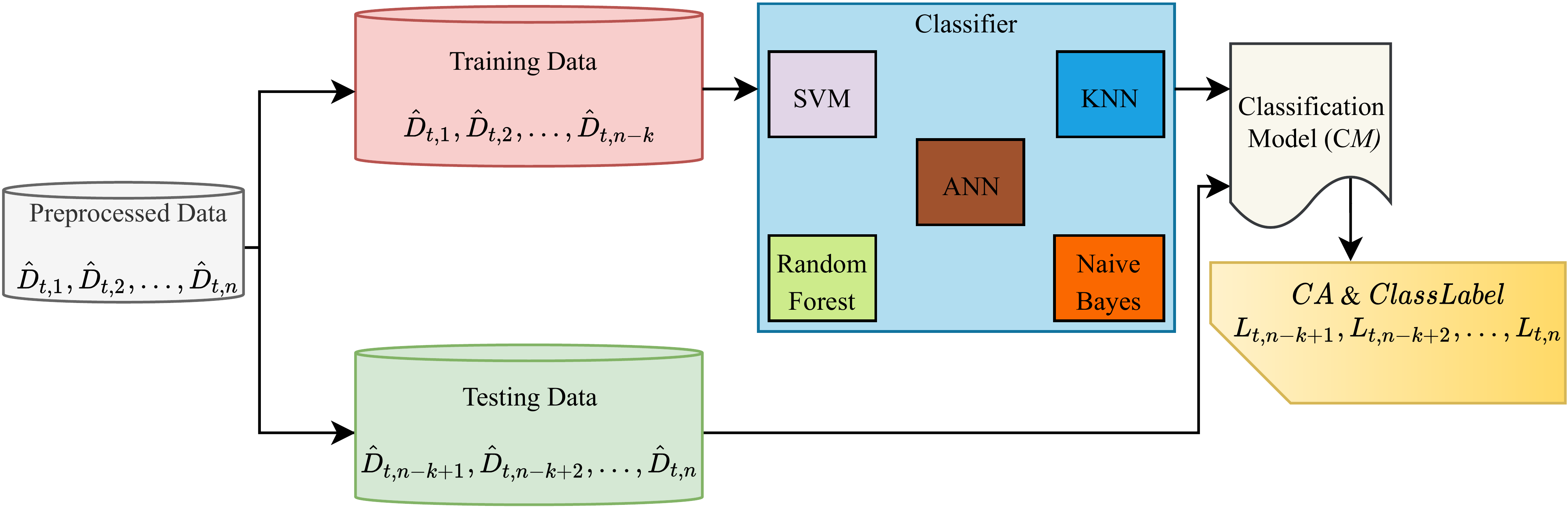}
		\caption{Classification flow for shared data}
		\label{fig:FC}
	\end{center}
\end{figure*}
Sanitized data $\hat{D}$ is the combination of both noisy data $\bar{D}^{S}$ and non-sensitive data $D^{NS}$, which is given to the input layer with $n$ nodes. A multi-layer perceptron (Fig. 4) architecture consists of three layers: one input layer, one hidden layer, and one output layer. The input layer receives the data and prepares it for feeding the hidden layer $H$ with $n_{2}$ nodes. The hidden layer $H$ is responsible to process the acquired results from input layer. The results of $H$ is $H_{n_{2}} = \{\varphi_{1}\hat{D}_{t_{1}}, \varphi_{2}\hat{D}_{t_{2}}, \dots, \varphi_{n_{2}}\hat{D}_{t_{n}}+b_{1}\}$,
where $\varphi$, $b_{1}$ are the weight and bias, respectively. The obtained results from the hidden layer is given to the output layer. The activation function is used to activate the neuron at the hidden layer as well as at the output layer. The results of the last layer with $n_{3}$ nodes as $y = \{\varphi_{1}h_{1}, \varphi_{2}h_{2}, \dots, \varphi_{n_{3}}h_{n_{2}}+b_{2}\}$ is achieved, where $b_{2}$ is the bias. The final classification result $L = \{L_{t,n-k+1},L_{t,n-k+2},\dots,L_{t,n}\}$ is obtained from \textit{$CM$}.
 	\begin{figure*}[!htbp]
	\begin{center}
	\includegraphics[width=0.95\textwidth]{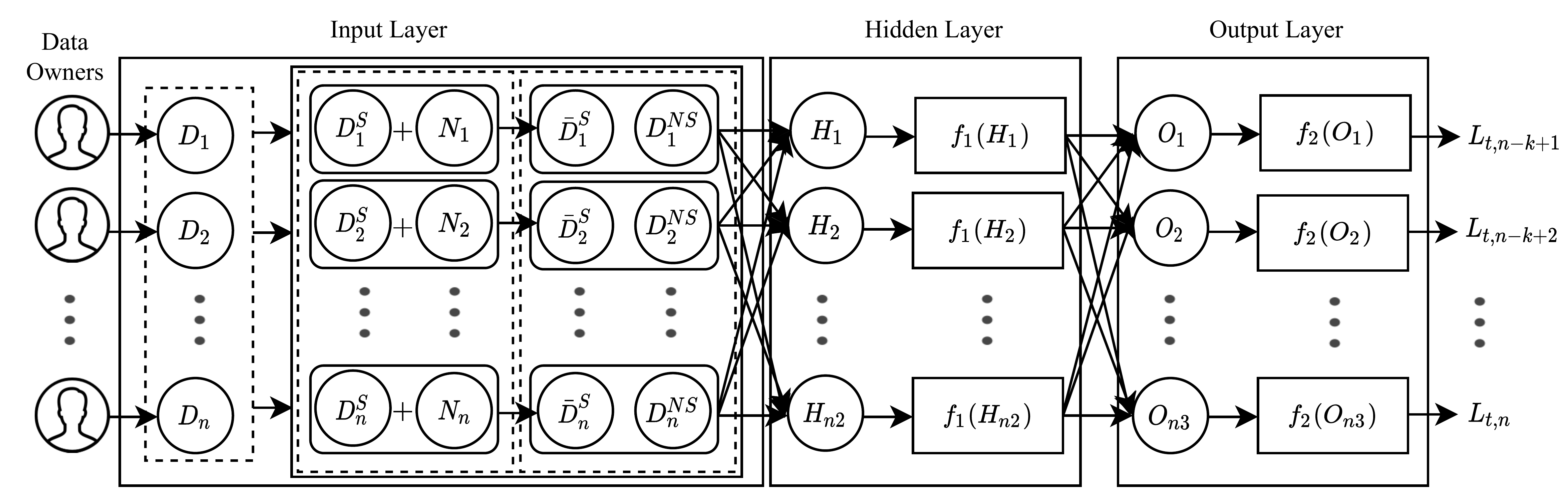}
	\caption{Learning Model for privacy preserving}
		\label{fig:NN}
	\end{center}
\end{figure*}
The steps for the data classification have been described by Algorithm 3. Steps 3 to 5 in this algorithm display the Support Vector Machine (SVM) classifier's efficiency. In steps 6 to 8, the procedure for the K-Nearest Neighbor (KNN) classifier is provided. Using steps 9 to 11, the Random Forest (RF) classifier classifies the data. The Naive Bayes (NB) classifier is conducted using step 12. Finally, the Artificial Neural Network (ANN) task is addressed in steps 13 to 18.
\par
In algorithm 1, steps 2 to 8 perform the classification over noised data using a classification model, whose time complexity depends on the data partition, noise addition, and classifier usage. The data is separated using step 3, which takes the time $\mathcal{O}(n^{2})$, and space $\mathcal{O}(n)$, where $n$ be the total number of input records. To preserve the privacy of the data, noise is generated in step 3 and added using the Laplace mechanism in step 4, which requires $\mathcal{O}(n^{2})$ time, and $\mathcal{O}(n)$ space. The classification model is performed using steps 7, which needs $\mathcal{O}(n^{3})$ time, and $\mathcal{O}(n^{2})$ space. Therefore, the total time complexity, and space complexity of  PPMD is $\mathcal{O}(n^{3})$ = ($\mathcal{O}(n^{2})$ + $\mathcal{O}(n^{2})$ + $\mathcal{O}(n^{3})$), $\mathcal{O}(n^{2})$ = ($\mathcal{O}(n)$ + $\mathcal{O}(n)$ + $\mathcal{O}(n^{2})$), respectively. PPMD complexity analysis implies that the aid of endurable time and space protects the data, which establishes its potency.
\begin{algorithm}
\KwIn{Input vector $\hat{D}$, weight $w$, bias $b$, activation function $f(x)$}
\KwOut{unknown class label $\bar{y}$, tree numbers $n\_tree$}
Initialize input vector $\hat{D}$ = \{$\hat{D}_{1}$, $\hat{D}_{2}$, $\dots$, $\hat{D}_{n}$\}, $\hat{D}_{1}$ = \{$(x_{1}, y_{1})$, $(x_{2}, y_{2})$, $\dots$, $(x_{\ddot{i}}, y_{\ddot{j}})$\}, $w$, $b$ \\
\textbf{for} $i$ = $1$, $2$, $\dots$, $n$ \textbf{do}\\
\hspace{.3cm}$z_{i}$ = $\sum$ $\hat{D}_{i}$ $\cdot$ $w^{T}$ + $b$ \\  
\hspace{.3cm}$f(z_{i})$ $>$ 0, $\bar{y}_{i}$ := 1 \\
\hspace{.3cm}$f(z_{i})$ $<$ 0, $\bar{y}_{i}$ := 0 \\
\hspace{.3cm}\textbf{for} Compute Set $I$ contains the minimum sets of $k$ \textbf{do}\\
\hspace{.6cm}distance $d$ ($\hat{D}_{i}$, $\bar{y}_{i}$) \\
\hspace{.3cm}\textbf{end for} \\
\hspace{.3cm}\textbf{for} $\ddot{t}$ = $1$, $2$, $\dots$, $n\_tree$  \textbf{do}\\
\hspace{.6cm} tree\_classification($\hat{D}_{i}$, $\bar{y}_{i}$)   \\
\hspace{.3cm}\textbf{end for} \\
\hspace{.3cm} $\bar{y}_{i}$ = $argmax_{y}$ $P(y)$ $\prod$ $P(X_{i}\mid y)$\\
\hspace{.3cm}\textbf{for} $\ddot{k}$ = $1$, $2$  \textbf{do}\\ 
\hspace{.6cm}\textbf{for} {$\ddot{l}$ = $1$, $\dots$, $n_{\ddot{k}+1}$}  \textbf{do} \\ 
\hspace{.9cm} $z_{\ddot{l}}^{(\ddot{k}+1)}$ = $\sum\limits_{\hat{t}=1}^{n}$ $x_{\hat{t}}^{(\ddot{k})}$ $\cdot$ $w_{\hat{t}}^{(\ddot{k})}$ + $b_{\ddot{l}}^{(\ddot{k})}$ \\  
\hspace{.9cm} $\bar{y}_{i}$ = $f(z_{\ddot{l}}^{(\ddot{k}+1)})$ \\
\hspace{.6cm}\textbf{end for} \\
\hspace{.3cm}\textbf{end for} \\
\hspace{.3cm}return $\bar{y}_{i}$ \\
\textbf{end for} 
\caption{Data Classification}
\label{algo:bjh}
\end{algorithm}
 \section{Performance evaluation}
 \subsection{Experimental setup}
The experiments are performed on a machine equipped with Intel (R) Core (TM) i5-4210U CPU @ 1.70GHz clock speed. The computing machine runs Ubuntu 64-bit and has 8 GB of the main memory RAM. Python 2.7.15 programming language is used to complete the classification tasks. The five distinct classifiers: SVM, KNN, RF, NB, and ANN, have been used to train \textit{CM} over training data. 
\subsection{Datasets and classification parameters}
Heart Disease, Arrhythmia, Hepatitis, Indian-liver-patient, and Framingham datasets are taken from the UCI Machine Learning Repository \cite{R20} to train \textit{CM}. There are 303, 452, 155, 583, 303 instances, 75, 280, 20, 11, 14 attributes, and 2, 13, 2, 2, 2 classes (binary and multi-class) in these datasets, correspondingly as shown in Table 7.
\begin{table}[!htbp]
\centering
		\caption{Basic information of four datasets}
		\label{TableExp236}
	\resizebox{\columnwidth}{!}{
			\begin{tabular}{|c|c|c|c|c|c|}\hline 
Dataset  & \#Instances & \#Features & \#Classes & Samples in  & Samples in  \\ 
&  &  &  &  training set &  test set\\ \hline
Heart Disease & 303 & 75 & 2 & 272 & 31\\ \hline
Arrhythmia & 452 & 280 & 13 & 406 & 46\\ \hline
Hepatitis & 155 & 20 & 2 & 139  & 16\\ \hline
Indian-liver-patient & 583 & 11 & 2 & 524 & 59\\ \hline 
Framingham & 303 & 14 & 2 & 272 & 31\\ \hline
\end{tabular}}
\end{table}
\par
To train \textit{CM}, 9/10 of data is used as training data from the entire dataset, while the rest is used as test data. For a particular case, there are 303 instances in the Heart Disease dataset. The 272 instances (i.e., 9/10 of 303 instances) are used as training samples and the remaining 31 instances for testing samples. The machine learning model is carried out over Clean, PPMD, PMLM \cite{R14}, NbAFL \cite{R29}, and MLPAM \cite{R19}. We have used the Laplace mechanism to generate the noise. However, PPMD, PMLM, NbAFL, and MLPAM schemes contain noise.
The results of \textit{$CM$} are measured using test data, and the \textit{$CA$}, \textit{$P$}, \textit{$R$}, and \textit{$FS$} are computed from these results. 
 \subsection{Results}
The \textit{$CM$} obtains the classification results including \textit{$CA$}, \textit{$P$}, \textit{$R$}, and \textit{$FS$} over Clean, PPMD, PMLM \cite{R14}, NbAFL \cite{R29}, and MLPAM \cite{R19}, as demonstrated in Figs. 5(a)-(e) to 8(a)-(e).
In PPMD, the maximum value of \textit{$CA$} is 93.75\% on the Arrhythmia dataset using the ANN classifier. The minimum value of \textit{$CA$} is 62.50\% on the Hepatitis dataset using the KNN classifier. The average value of \textit{$CA$} is 72.20\%, 73.40\%, 73.61\%, 70.03\%, and 84.11\% over Heart Disease, Arrhythmia, Hepatitis, Indian-liver-patient, and Framingham dataset, respectively.
The highest value of \textit{$P$} is 94.11\% on the Indian-liver-patient dataset using the NB classifier. The lowest value of \textit{$P$} is 43.33\% on the Heart Disease dataset using the ANN classifier. The average value of \textit{$P$} is 69.33\%, 53.51\%, 74.18\%, 78.23\%, and 76.15\% over Heart Disease, Arrhythmia, Hepatitis, Indian-liver-patient, and Framingham dataset, respectively.
The maximum value of \textit{$R$} is 100\% on the Indian-liver-patient dataset using the SVM classifier. The minimum value of \textit{$R$} is 39.13\% on the Arrhythmia dataset using the ANN classifier. The average value of \textit{$R$} is 59.62\%, 62.48\%, 82.30\%, 76.95\%, and 84.06\% over Heart Disease, Arrhythmia, Hepatitis, Indian-liver-patient, and Framingham dataset, respectively.
The highest value of \textit{$FS$} is 87.99\% on the Hepatitis dataset using the RF classifier. The lowest value of \textit{$FS$} is 41.37\% on the Arrhythmia dataset using the ANN classifier. The average value of \textit{$FS$} is 62.84\%, 57.20\%, 79.48\%, 75.97\%, and 78.77\% over Heart Disease, Arrhythmia, Hepatitis, Indian-liver-patient, and Framingham dataset, respectively.
The datasets’ performance descends in order: Framingham, Hepatitis, Indian-liver-patient, Heart Disease, and Arrhythmia.
\begin{figure}[!htbp]
\centering
\begin{subfigure}[t]{0.49\textwidth}
\begin{tikzpicture}[node distance = 1cm,auto,scale=.70, transform shape]
\pgfplotsset{every axis y label/.append style={rotate=180,yshift=10.5cm}}
\begin{axis}[
       axis on top=false,
       xmin=12, xmax=110,
       ymin=0, ymax=1.1,
       xtick={22,42,61,82,100},
       xticklabels={HD,Arr,Hep,IndLiv,Fram},
        ycomb,
        ylabel near ticks, yticklabel pos=left,
       ylabel={CA},
       legend style={at={(0.5,-0.15)},
       anchor=north,legend columns=3},
       ymajorgrids=true,
       grid style=dashed,
          ]
\addplot+[mark options={fill=blue},fill=blue!40!,draw=blue,  thick]
coordinates
{(17,.8064) (36,.6956) (55,.7500) (74,.7627) (93,.8514)}
\closedcycle;%
\addlegendentry{Clean Data}

\addplot+[mark options={fill=green},fill=pink,draw=green,  thick] 
coordinates
 {(20,.7741) (39,.6739) (58,.6875) (77,.7457) (96,.8443) }
\closedcycle;%
\addlegendentry{PPMD}

\addplot+[mark options={fill=white},fill=red!60!,draw=red!70!,  thick] 
coordinates
 {(23,.7419) (42,.6521) (61,.6250) (80,.7288) (99,.8372)}
\closedcycle;%
\addlegendentry{PMLM \cite{R14}}
\addplot+[mark options={fill=Mustard},fill=Mustard,draw=Mustard,  thick] 
coordinates
{(26,.7741) (45,.6521) (64,.6250) (83,.7118) (102,.8419) }
\closedcycle;%
\addlegendentry{NbAFL \cite{R29}}
\addplot+[mark options={fill=teal},fill=teal,draw=teal,  thick] 
coordinates
 {(29,.7096) (48,.6304) (67,.6250) (86,.6949) (105,.8231)}
\closedcycle;%
\addlegendentry{MLPAM \cite{R19}}
\end{axis}
\end{tikzpicture}
            \caption{SVM}
        \end{subfigure}
                    \hfill
\begin{subfigure}[t]{0.49\textwidth}
\begin{tikzpicture}[node distance = 1cm,auto,scale=.70, transform shape]
\pgfplotsset{every axis y label/.append style={rotate=180,yshift=10.5cm}}
\begin{axis}[
      axis on top=false,
       xmin=12, xmax=110,
       ymin=0, ymax=1.1,
       xtick={22,42,61,82,100},
       xticklabels={HD,Arr,Hep,IndLiv,Fram},
        ycomb,
        ylabel near ticks, yticklabel pos=left,
       ylabel={CA},
       legend style={at={(0.5,-0.15)},
       anchor=north,legend columns=3},
       ymajorgrids=true,
       grid style=dashed,
          ]
\addplot+[mark options={fill=blue},fill=blue!40!,draw=blue,  thick]
coordinates
{(17,.7741) (36,.7391) (55,.6875) (74,.7288) (93,.8278)}
\closedcycle;%
\addlegendentry{Clean Data}

\addplot+[mark options={fill=green},fill=pink,draw=green,  thick] 
coordinates
 {(20,.7419) (39,.7173) (58,.6250) (77,.7118) (96,.8207) }
\closedcycle;%
\addlegendentry{PPMD}

\addplot+[mark options={fill=white},fill=red!60!,draw=red!70!,  thick] 
coordinates
{(23,.7096) (42,.6521) (61,.5625) (80,.6949) (99,.8136)}
\closedcycle;%
\addlegendentry{PMLM \cite{R14}}
\addplot+[mark options={fill=Mustard},fill=Mustard,draw=Mustard,  thick] 
coordinates
{(26,.7096) (45,.6956) (64,.6250) (83,.6949) (102,.8183) }
\closedcycle;%
\addlegendentry{NbAFL \cite{R29}}
\addplot+[mark options={fill=teal},fill=teal,draw=teal,  thick] 
coordinates
 {(29,.7096) (48,.6521) (67,.6250) (86,.6779) (105,.8089)}
\closedcycle;%
\addlegendentry{MLPAM \cite{R19}}
\end{axis}
\end{tikzpicture}
            \caption{KNN}
        \end{subfigure}
\begin{subfigure}[t]{0.49\textwidth}
\begin{tikzpicture}[node distance = 1cm,auto,scale=.70, transform shape]
\pgfplotsset{every axis y label/.append style={rotate=180,yshift=10.5cm}}
\begin{axis}[
      axis on top=false,
       xmin=12, xmax=110,
       ymin=0, ymax=1.1,
       xtick={22,42,61,82,100},
       xticklabels={HD,Arr,Hep,IndLiv,Fram},
        ycomb,
        ylabel near ticks, yticklabel pos=left,
       ylabel={CA},
       legend style={at={(0.5,-0.15)},
       anchor=north,legend columns=3},
       ymajorgrids=true,
       grid style=dashed,
          ]
\addplot+[mark options={fill=blue},fill=blue!40!,draw=blue,  thick]
coordinates
{(17,.7096) (36,.7173) (55,.8750) (74,.6610) (93,.8272)}
\closedcycle;%
\addlegendentry{Clean Data}

\addplot+[mark options={fill=green},fill=pink,draw=green,  thick] 
coordinates
 {(20,.6875) (39,.6956) (58,.8125) (77,.6440) (96,.8231) }
\closedcycle;%
\addlegendentry{PPMD}

\addplot+[mark options={fill=white},fill=red!60!,draw=red!70!,  thick] 
coordinates
 {(23,.6774) (42,.6521) (61,.6875) (80,.6271) (99,.8113)}
\closedcycle;%
\addlegendentry{PMLM \cite{R14}}
\addplot+[mark options={fill=Mustard},fill=Mustard,draw=Mustard,  thick] 
coordinates
 {(26,.6774) (45,.6739) (64,.7500) (83,.6101) (102,.8160) }
\closedcycle;%
\addlegendentry{NbAFL \cite{R29}}
\addplot+[mark options={fill=teal},fill=teal,draw=teal,  thick] 
coordinates
 {(29,.6774) (48,.6086) (67,.6875) (86,.5932) (105,.8018)}
\closedcycle;%
\addlegendentry{MLPAM \cite{R19}}
\end{axis}
\end{tikzpicture}
            \caption{RF}
        \end{subfigure}
                    \hfill
\begin{subfigure}[t]{0.49\textwidth}
\begin{tikzpicture}[node distance = 1cm,auto,scale=.70, transform shape]
\pgfplotsset{every axis y label/.append style={rotate=180,yshift=10.5cm}}
\begin{axis}[
      axis on top=false,
       xmin=12, xmax=110,
       ymin=0, ymax=1.1,
       xtick={22,42,61,82,100},
       xticklabels={HD,Arr,Hep,IndLiv,Fram},
        ycomb,
        ylabel near ticks, yticklabel pos=left,
       ylabel={CA},
       legend style={at={(0.5,-0.15)},
       anchor=north,legend columns=3},
       ymajorgrids=true,
       grid style=dashed,
          ]
\addplot+[mark options={fill=blue},fill=blue!40!,draw=blue,  thick]
coordinates
{(17,.7419) (36,.6635) (55,.8125) (74,.6440) (93,.8490)}
\closedcycle;%
\addlegendentry{Clean Data}

\addplot+[mark options={fill=green},fill=pink,draw=green,  thick] 
coordinates
 {(20,.7096) (39,.6460) (58,.7500) (77,.6271) (96,.8349) }
\closedcycle;%
\addlegendentry{PPMD}

\addplot+[mark options={fill=white},fill=red!60!,draw=red!70!,  thick] 
coordinates
{(23,.6774) (42,.6304) (61,.6875) (80,.6101) (99,.8254)}
\closedcycle;%
\addlegendentry{PMLM \cite{R14}}
\addplot+[mark options={fill=Mustard},fill=Mustard,draw=Mustard,  thick] 
coordinates
 {(26,.6774) (45,.6304) (64,.6875) (83,.6101) (102,.8325) }
\closedcycle;%
\addlegendentry{NbAFL \cite{R29}}
\addplot+[mark options={fill=teal},fill=teal,draw=teal,  thick] 
coordinates
 {(29,.6721) (48,.6086) (67,.6875) (86,.5932) (105,.8136)}
\closedcycle;%
\addlegendentry{MLPAM \cite{R19}}
\end{axis}
\end{tikzpicture}
            \caption{NB}
        \end{subfigure}
                    \hfill
\begin{subfigure}[t]{0.49\textwidth}
\begin{tikzpicture}[node distance = 1cm,auto,scale=.70, transform shape]
\pgfplotsset{every axis y label/.append style={rotate=180,yshift=10.5cm}}
\begin{axis}[
      axis on top=false,
       xmin=12, xmax=110,
       ymin=0, ymax=1.1,
       xtick={22,42,61,82,100},
       xticklabels={HD,Arr,Hep,IndLiv,Fram},
        ycomb,
        ylabel near ticks, yticklabel pos=left,
       ylabel={CA},
       legend style={at={(0.5,-0.15)},
       anchor=north,legend columns=3},
       ymajorgrids=true,
       grid style=dashed,
          ]
\addplot+[mark options={fill=blue},fill=blue!40!,draw=blue,  thick]
coordinates
{(17,.7034) (36,.9440) (55,.8321) (74,.8034) (93,.8837)}
\closedcycle;%
\addlegendentry{Clean Data}

\addplot+[mark options={fill=green},fill=pink,draw=green,  thick] 
coordinates
 {(20,.6973) (39,.9375) (58,.8058) (77,.7729) (96,.8825) }
\closedcycle;%
\addlegendentry{PPMD}

\addplot+[mark options={fill=white},fill=red!60!,draw=red!70!,  thick] 
coordinates
 {(23,.6912) (42,.9350) (61,.6547) (80,.7424) (99,.8801)}
\closedcycle;%
\addlegendentry{PMLM \cite{R14}}
\addplot+[mark options={fill=Mustard},fill=Mustard,draw=Mustard,  thick] 
coordinates
 {(26,.6667) (45,.9350) (64,.7258) (83,.7583) (102,.8712) }
\closedcycle;%
\addlegendentry{NbAFL \cite{R29}}
\addplot+[mark options={fill=teal},fill=teal,draw=teal,  thick] 
coordinates
 {(29,.6593) (48,.9267) (67,.6475) (86,.7405) (105,.7905)}
\closedcycle;%
\addlegendentry{MLPAM \cite{R19}}
\end{axis}
\end{tikzpicture}
            \caption{ANN}
        \end{subfigure}
\caption{Accuracy of \textit{CM} in PPMD}
    \end{figure}
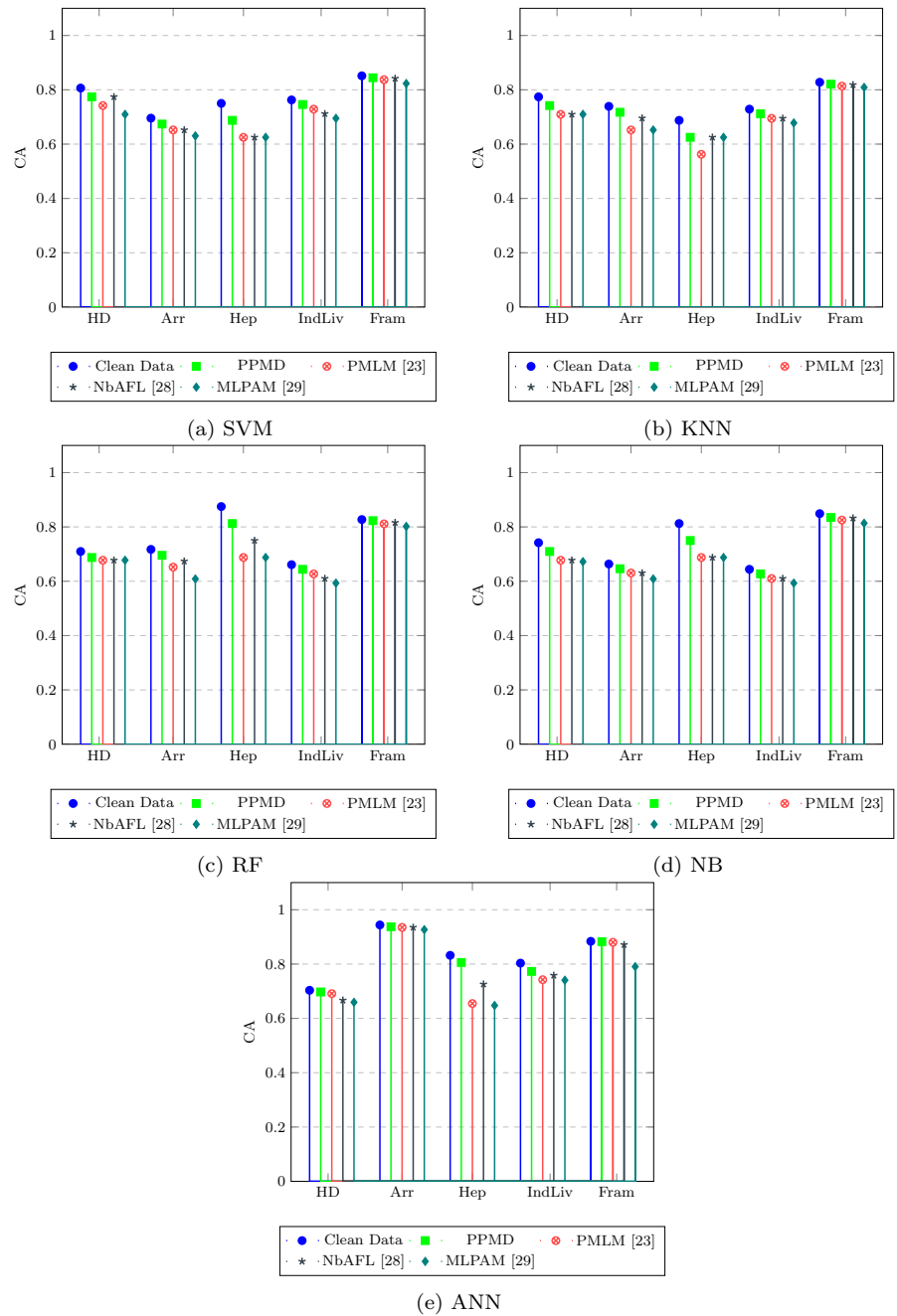                  

\begin{figure}[!htbp]
\centering
\begin{subfigure}[t]{0.49\textwidth}
\begin{tikzpicture}[node distance = 1cm,auto,scale=.70, transform shape]
\pgfplotsset{every axis y label/.append style={rotate=180,yshift=10.5cm}}
\begin{axis}[
       axis on top=false,
       xmin=12, xmax=110,
       ymin=0, ymax=1.1,
       xtick={22,42,61,82,100},
       xticklabels={HD,Arr,Hep,IndLiv,Fram},
        ycomb,
        ylabel near ticks, yticklabel pos=left,
       ylabel={\textit{$P$}},
       legend style={at={(0.5,-0.15)},
       anchor=north,legend columns=3},
       ymajorgrids=true,
       grid style=dashed,
          ]
\addplot+[mark options={fill=blue},fill=blue!40!,draw=blue,  thick]
coordinates
{(17,.7692) (36,.5146) (55,.7857) (74,.7627) (93,.7323)}
\closedcycle;%
\addlegendentry{Clean Data}

\addplot+[mark options={fill=green},fill=pink,draw=green,  thick] 
coordinates
 {(20,.7500) (39,.4839) (58,.7333) (77,.7457) (96,.7286) }
\closedcycle;%
\addlegendentry{PPMD}

\addplot+[mark options={fill=white},fill=red!60!,draw=red!70!,  thick] 
coordinates
 {(23,.7000) (42,.4253) (61,.6923) (80,.7288) (99,.7010) }
\closedcycle;%
\addlegendentry{PMLM \cite{R14}}
\addplot+[mark options={fill=Mustard},fill=Mustard,draw=Mustard,  thick] 
coordinates
 {(26,.7096) (45,.4281) (64,.6875) (83,.7118) (102,.7168) }
\closedcycle;%
\addlegendentry{NbAFL \cite{R29}}
\addplot+[mark options={fill=teal},fill=teal,draw=teal,  thick] 
coordinates
 {(29,.6774) (48,.4246) (67,.6875) (86,.6842) (105,.6877)}
\closedcycle;%
\addlegendentry{MLPAM \cite{R19}}
\end{axis}
\end{tikzpicture}
            \caption{SVM}
        \end{subfigure}
            \hfill
\begin{subfigure}[t]{0.49\textwidth}
\begin{tikzpicture}[node distance = 1cm,auto,scale=.70, transform shape]
\pgfplotsset{every axis y label/.append style={rotate=180,yshift=10.5cm}}
\begin{axis}[
        axis on top=false,
       xmin=12, xmax=110,
       ymin=0, ymax=1.1,
       xtick={22,42,61,82,100},
       xticklabels={HD,Arr,Hep,IndLiv,Fram},
        ycomb,
        ylabel near ticks, yticklabel pos=left,
       ylabel={\textit{$P$}},
       legend style={at={(0.5,-0.15)},
       anchor=north,legend columns=3},
       ymajorgrids=true,
       grid style=dashed,
          ]
\addplot+[mark options={fill=blue},fill=blue!40!,draw=blue,  thick]
coordinates
{(17,.8000) (36,.6643) (55,.7142) (74,.8125) (93,.7578) }
\closedcycle;%
\addlegendentry{Clean Data}

\addplot+[mark options={fill=green},fill=pink,draw=green,  thick] 
coordinates
 {(20,.7692) (39,.6415) (58,.7142) (77,.7826) (96,.7483) }
\closedcycle;%
\addlegendentry{PPMD}

\addplot+[mark options={fill=white},fill=red!60!,draw=red!70!,  thick] 
coordinates
{(23,.7500) (42,.5485) (61,.6666) (80,.7659) (99,.7393) }
\closedcycle;%
\addlegendentry{PMLM \cite{R14}}
\addplot+[mark options={fill=Mustard},fill=Mustard,draw=Mustard,  thick] 
coordinates
{(26,.7475) (45,.5570) (64,.6923) (83,.7551) (102,.7409) }
\closedcycle;%
\addlegendentry{NbAFL \cite{R29}}
\addplot+[mark options={fill=teal},fill=teal,draw=teal,  thick] 
coordinates
 {(29,.7111) (48,.5338) (67,.6923) (86,.7500) (105,.7254)}
\closedcycle;%
\addlegendentry{MLPAM \cite{R19}}
\end{axis}
\end{tikzpicture}
            \caption{KNN}
        \end{subfigure}
            \hfill
\begin{subfigure}[t]{0.49\textwidth}
\begin{tikzpicture}[node distance = 1cm,auto,scale=.70, transform shape]
\pgfplotsset{every axis y label/.append style={rotate=180,yshift=10.5cm}}
\begin{axis}[
       axis on top=false,
       xmin=12, xmax=110,
       ymin=0, ymax=1.1,
       xtick={22,42,61,82,100},
       xticklabels={HD,Arr,Hep,IndLiv,Fram},
        ycomb,
        ylabel near ticks, yticklabel pos=left,
       ylabel={\textit{$P$}},
       legend style={at={(0.5,-0.15)},
       anchor=north,legend columns=3},
       ymajorgrids=true,
       grid style=dashed,
          ]
\addplot+[mark options={fill=blue},fill=blue!40!,draw=blue,  thick]
coordinates
{(17,.8461) (36,.6006) (55,.8666) (74,.7142) (93,.7752) }
\closedcycle;%
\addlegendentry{Clean Data}

\addplot+[mark options={fill=green},fill=pink,draw=green,  thick] 
coordinates
 {(20,.8000) (39,.5626) (58,.7857) (77,.7045) (96,.7731) }
\closedcycle;%
\addlegendentry{PPMD}

\addplot+[mark options={fill=white},fill=red!60!,draw=red!70!,  thick] 
coordinates
 {(23,.5555) (42,.5232) (61,.7692) (80,.6444) (99,.7527) }
\closedcycle;%
\addlegendentry{PMLM \cite{R14}}
\addplot+[mark options={fill=Mustard},fill=Mustard,draw=Mustard,  thick] 
coordinates
{(26,.5706) (45,.5533) (64,.7692) (83,.7021) (102,.7554) }
\closedcycle;%
\addlegendentry{NbAFL \cite{R29}}
\addplot+[mark options={fill=teal},fill=teal,draw=teal,  thick] 
coordinates
 {(29,.5161) (48,.5126) (67,.6666) (86,.6585) (105,.7407)}
\closedcycle;%
\addlegendentry{MLPAM \cite{R19}}
\end{axis}
\end{tikzpicture}
            \caption{RF}
        \end{subfigure}
            \hfill
\begin{subfigure}[t]{0.49\textwidth}
\begin{tikzpicture}[node distance = 1cm,auto,scale=.70, transform shape]
\pgfplotsset{every axis y label/.append style={rotate=180,yshift=10.5cm}}
\begin{axis}[
         axis on top=false,
       xmin=12, xmax=110,
       ymin=0, ymax=1.1,
       xtick={22,42,61,82,100},
       xticklabels={HD,Arr,Hep,IndLiv,Fram},
        ycomb,
        ylabel near ticks, yticklabel pos=left,
       ylabel={\textit{$P$}},
       legend style={at={(0.5,-0.15)},
       anchor=north,legend columns=3},
       ymajorgrids=true,
       grid style=dashed,
          ]
\addplot+[mark options={fill=blue},fill=blue!40!,draw=blue,  thick]
coordinates
{(17,.7500) (36,.5869) (55,.9000) (74,.9444) (93,.8245) }
\closedcycle;%
\addlegendentry{Clean Data}

\addplot+[mark options={fill=green},fill=pink,draw=green,  thick] 
coordinates
 {(20,.7142) (39,.5485) (58,.8333) (77,.9411) (96,.8128) }
\closedcycle;%
\addlegendentry{PPMD}

\addplot+[mark options={fill=white},fill=red!60!,draw=red!70!,  thick] 
coordinates
 {(23,.6875) (42,.5372) (61,.7692) (80,.9166) (99,.7966) }
\closedcycle;%
\addlegendentry{PMLM \cite{R14}}
\addplot+[mark options={fill=Mustard},fill=Mustard,draw=Mustard,  thick] 
coordinates
 {(26,.7029) (45,.5359) (64,.7857) (83,.9230) (102,.8100) }
\closedcycle;%
\addlegendentry{NbAFL \cite{R29}}
\addplot+[mark options={fill=teal},fill=teal,draw=teal,  thick] 
coordinates
 {(29,.6684) (48,.5309) (67,.7136) (86,.8888) (105,.7902)}
\closedcycle;%
\addlegendentry{MLPAM \cite{R19}}
\end{axis}
\end{tikzpicture}
            \caption{NB}
        \end{subfigure}
            \hfill
\begin{subfigure}[t]{0.49\textwidth}
\begin{tikzpicture}[node distance = 1cm,auto,scale=.70, transform shape]
\pgfplotsset{every axis y label/.append style={rotate=180,yshift=10.5cm}}
\begin{axis}[
       axis on top=false,
       xmin=12, xmax=110,
       ymin=0, ymax=1.1,
       xtick={22,42,61,82,100},
       xticklabels={HD,Arr,Hep,IndLiv,Fram},
        ycomb,
        ylabel near ticks, yticklabel pos=left,
       ylabel={\textit{$P$}},
       legend style={at={(0.5,-0.15)},
       anchor=north,legend columns=3},
       ymajorgrids=true,
       grid style=dashed,
          ]
\addplot+[mark options={fill=blue},fill=blue!40!,draw=blue,  thick]
coordinates
{(17,.4516) (36,.4500) (55,.7222) (74,.7500) (93,.7453) }
\closedcycle;%
\addlegendentry{Clean Data}

\addplot+[mark options={fill=green},fill=pink,draw=green,  thick] 
coordinates
 {(20,.4333) (39,.4390) (58,.6428) (77,.7377) (96,.7451) }
\closedcycle;%
\addlegendentry{PPMD}

\addplot+[mark options={fill=white},fill=red!60!,draw=red!70!,  thick] 
coordinates
 {(23,.4285) (42,.4390) (61,.6000) (80,.6428) (99,.7442) }
\closedcycle;%
\addlegendentry{PMLM \cite{R14}}
\addplot+[mark options={fill=Mustard},fill=Mustard,draw=Mustard,  thick] 
coordinates
 {(26,.4285) (45,.4337) (64,.6180) (83,.7000) (102,.7343) }
\closedcycle;%
\addlegendentry{NbAFL \cite{R29}}
\addplot+[mark options={fill=teal},fill=teal,draw=teal,  thick] 
coordinates
 {(29,.4285) (48,.4285) (67,.5833) (86,.6428) (105,.7322)}
\closedcycle;%
\addlegendentry{MLPAM \cite{R19}}
\end{axis}
\end{tikzpicture}
            \caption{ANN}
        \end{subfigure}
\caption{Precision of \textit{CM} in PPMD}
    \end{figure}
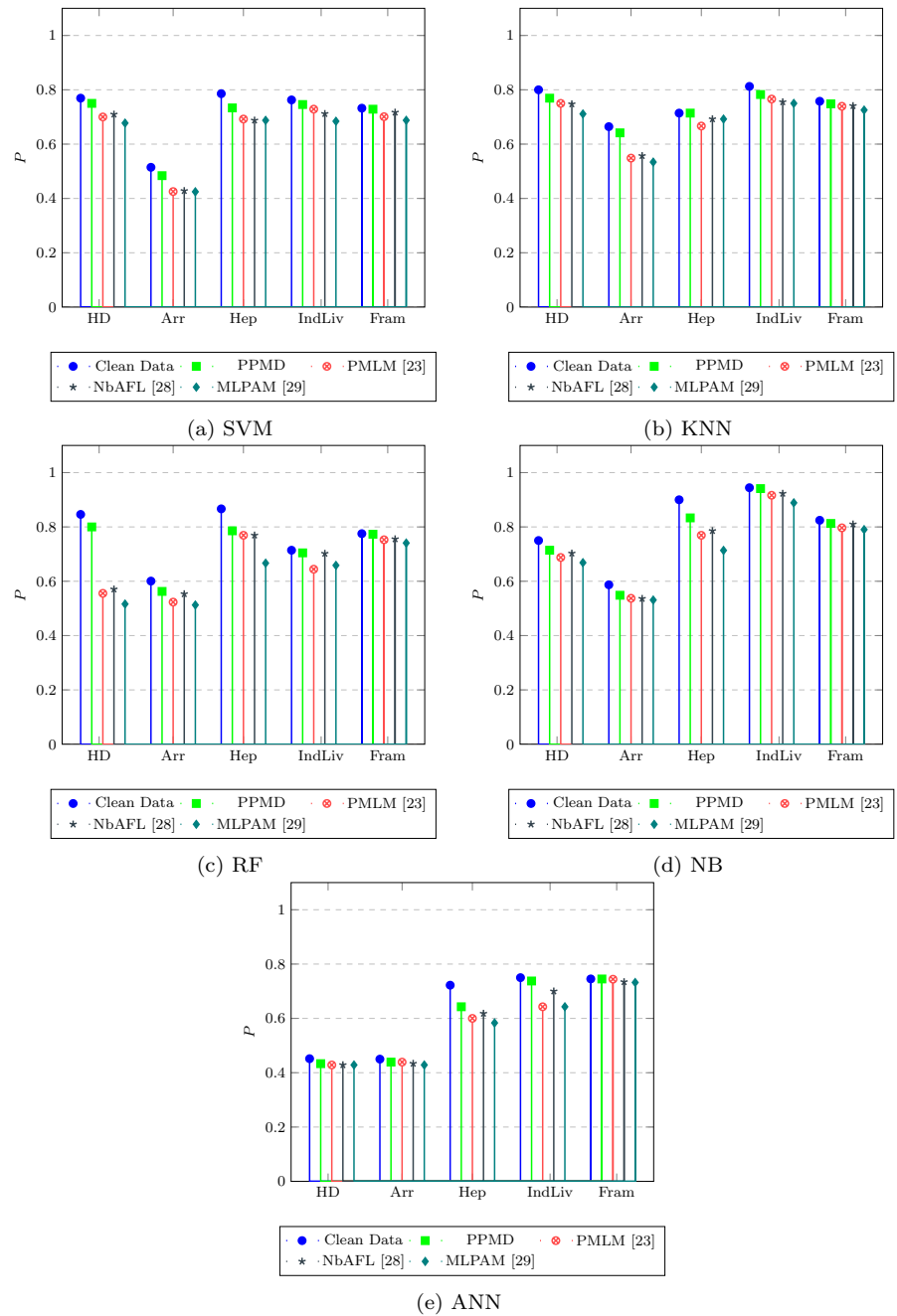
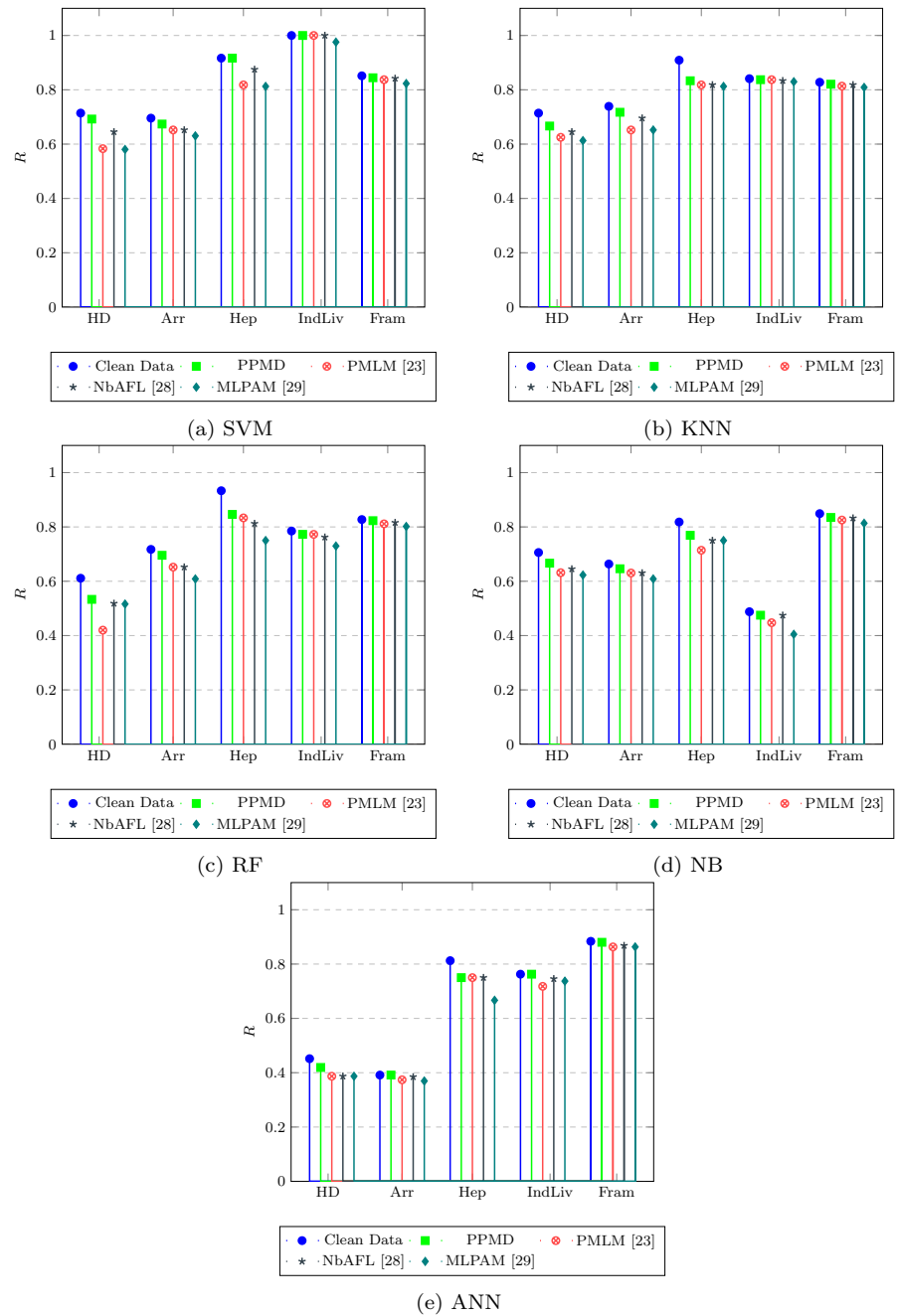
\begin{figure}[!htbp]
\centering
\begin{subfigure}[t]{0.49\textwidth}
\begin{tikzpicture}[node distance = 1cm,auto,scale=.70, transform shape]
\pgfplotsset{every axis y label/.append style={rotate=180,yshift=10.5cm}}
\begin{axis}[
     axis on top=false,
       xmin=12, xmax=110,
       ymin=0, ymax=1.1,
       xtick={22,42,61,82,100},
      xticklabels={HD,Arr,Hep,IndLiv,Fram},
        ycomb,
        ylabel near ticks, yticklabel pos=left,
      ylabel={\textit{$R$}},
      legend style={at={(0.5,-0.15)},
      anchor=north,legend columns=3},
      ymajorgrids=true,
      grid style=dashed,
          ]
\addplot+[mark options={fill=blue},fill=blue!40!,draw=blue,  thick]
coordinates
{(17,.7142) (36,.6956) (55,.9166) (74,1) (93,.8514) }
\closedcycle;%
\addlegendentry{Clean Data}

\addplot+[mark options={fill=green},fill=pink,draw=green,  thick] 
coordinates
 {(20,.6923) (39,.6739) (58,.9166) (77,1) (96,.8443) }
\closedcycle;%
\addlegendentry{PPMD}

\addplot+[mark options={fill=white},fill=red!60!,draw=red!70!,  thick] 
coordinates
{(23,.5833) (42,.6521) (61,.8181) (80,1) (99,.8372) }
\closedcycle;%
\addlegendentry{PMLM \cite{R14}}
\addplot+[mark options={fill=Mustard},fill=Mustard,draw=Mustard,  thick] 
coordinates
 {(26,.6451) (45,.6521) (64,.8750) (83,1) (102,.8419) }
\closedcycle;%
\addlegendentry{NbAFL \cite{R29}}
\addplot+[mark options={fill=teal},fill=teal,draw=teal,  thick] 
coordinates
 {(29,.5804) (48,.6304) (67,.8125) (86,.9756) (105,.8231)}
\closedcycle;%
\addlegendentry{MLPAM \cite{R19}}
\end{axis}
\end{tikzpicture}
            \caption{SVM}
        \end{subfigure}
            \hfill
\begin{subfigure}[t]{0.49\textwidth}
\begin{tikzpicture}[node distance = 1cm,auto,scale=.70, transform shape]
\pgfplotsset{every axis y label/.append style={rotate=180,yshift=10.5cm}}
\begin{axis}[
     axis on top=false,
       xmin=12, xmax=110,
       ymin=0, ymax=1.1,
       xtick={22,42,61,82,100},
      xticklabels={HD,Arr,Hep,IndLiv,Fram},
        ycomb,
        ylabel near ticks, yticklabel pos=left,
      ylabel={\textit{$R$}},
      legend style={at={(0.5,-0.15)},
      anchor=north,legend columns=3},
      ymajorgrids=true,
      grid style=dashed,
          ]
\addplot+[mark options={fill=blue},fill=blue!40!,draw=blue,  thick]
coordinates
{(17,.7142) (36,.7391) (55,.9090) (74,.8409) (93,.8278) }
\closedcycle;%
\addlegendentry{Clean Data}

\addplot+[mark options={fill=green},fill=pink,draw=green,  thick] 
coordinates
 {(20,.6666) (39,.7173) (58,.8333) (77,.8372) (96,.8207) }
\closedcycle;%
\addlegendentry{PPMD}

\addplot+[mark options={fill=white},fill=red!60!,draw=red!70!,  thick] 
coordinates
 {(23,.6250) (42,.6521) (61,.8181) (80,.8372) (99,.8136) }
\closedcycle;%
\addlegendentry{PMLM \cite{R14}}
\addplot+[mark options={fill=Mustard},fill=Mustard,draw=Mustard,  thick] 
coordinates
 {(26,.6451) (45,.6956) (64,.8181) (83,.8333) (102,.8183) }
\closedcycle;%
\addlegendentry{NbAFL \cite{R29}}
\addplot+[mark options={fill=teal},fill=teal,draw=teal,  thick] 
coordinates
 {(29,.6129) (48,.6521) (67,.8125) (86,.8297) (105,.8089)}
\closedcycle;%
\addlegendentry{MLPAM \cite{R19}}
\end{axis}
\end{tikzpicture}
            \caption{KNN}
        \end{subfigure}
            \hfill
\begin{subfigure}[t]{0.49\textwidth}
\begin{tikzpicture}[node distance = 1cm,auto,scale=.70, transform shape]
\pgfplotsset{every axis y label/.append style={rotate=180,yshift=10.5cm}}
\begin{axis}[
     axis on top=false,
       xmin=12, xmax=110,
       ymin=0, ymax=1.1,
       xtick={22,42,61,82,100},
      xticklabels={HD,Arr,Hep,IndLiv,Fram},
        ycomb,
        ylabel near ticks, yticklabel pos=left,
      ylabel={\textit{$R$}},
      legend style={at={(0.5,-0.15)},
      anchor=north,legend columns=3},
      ymajorgrids=true,
      grid style=dashed,
          ]
\addplot+[mark options={fill=blue},fill=blue!40!,draw=blue,  thick]
coordinates
{(17,.6111) (36,.7173) (55,.9333) (74,.7849) (93,.8272) }
\closedcycle;%
\addlegendentry{Clean Data}

\addplot+[mark options={fill=green},fill=pink,draw=green,  thick] 
coordinates
 {(20,.5333) (39,.6956) (58,.8461) (77,.7727) (96,.8231) }
\closedcycle;%
\addlegendentry{PPMD}

\addplot+[mark options={fill=white},fill=red!60!,draw=red!70!,  thick] 
coordinates
 {(23,.4200) (42,.6521) (61,.8333) (80,.7727) (99,.8113) }
\closedcycle;%
\addlegendentry{PMLM \cite{R14}}
\addplot+[mark options={fill=Mustard},fill=Mustard,draw=Mustard,  thick] 
coordinates
 {(26,.5183) (45,.6521) (64,.8125) (83,.7621) (102,.8160) }
\closedcycle;%
\addlegendentry{NbAFL \cite{R29}}
\addplot+[mark options={fill=teal},fill=teal,draw=teal,  thick] 
coordinates
 {(29,.5161) (48,.6086) (67,.7500) (86,.7297) (105,.8018)}
\closedcycle;%
\addlegendentry{MLPAM \cite{R19}}
\end{axis}
\end{tikzpicture}
            \caption{RF}
        \end{subfigure}
            \hfill
\begin{subfigure}[t]{0.49\textwidth}
\begin{tikzpicture}[node distance = 1cm,auto,scale=.70, transform shape]
\pgfplotsset{every axis y label/.append style={rotate=180,yshift=10.5cm}}
\begin{axis}[
     axis on top=false,
       xmin=12, xmax=110,
       ymin=0, ymax=1.1,
       xtick={22,42,61,82,100},
      xticklabels={HD,Arr,Hep,IndLiv,Fram},
        ycomb,
        ylabel near ticks, yticklabel pos=left,
      ylabel={\textit{$R$}},
      legend style={at={(0.5,-0.15)},
      anchor=north,legend columns=3},
      ymajorgrids=true,
      grid style=dashed,
          ]
\addplot+[mark options={fill=blue},fill=blue!40!,draw=blue,  thick]
coordinates
{(17,.7058) (36,.6635) (55,.8181) (74,.4878) (93,.8490) }
\closedcycle;%
\addlegendentry{Clean Data}

\addplot+[mark options={fill=green},fill=pink,draw=green,  thick] 
coordinates
 {(20,.6666) (39,.6460) (58,.7692) (77,.4750) (96,.8349) }
\closedcycle;%
\addlegendentry{PPMD}

\addplot+[mark options={fill=white},fill=red!60!,draw=red!70!,  thick] 
coordinates
{(23,.6315) (42,.6304) (61,.7142) (80,.4473) (99,.8254) }
\closedcycle;%
\addlegendentry{PMLM \cite{R14}}
\addplot+[mark options={fill=Mustard},fill=Mustard,draw=Mustard,  thick] 
coordinates
{(26,.6451) (45,.6304) (64,.7500) (83,.4745) (102,.8325) }
\closedcycle;%
\addlegendentry{NbAFL \cite{R29}}
\addplot+[mark options={fill=teal},fill=teal,draw=teal,  thick] 
coordinates
 {(29,.6229) (48,.6086) (67,.7500) (86,.4047) (105,.8136)}
\closedcycle;%
\addlegendentry{MLPAM \cite{R19}}
\end{axis}
\end{tikzpicture}
            \caption{NB}
        \end{subfigure}
            \hfill
\begin{subfigure}[t]{0.49\textwidth}
\begin{tikzpicture}[node distance = 1cm,auto,scale=.70, transform shape]
\pgfplotsset{every axis y label/.append style={rotate=180,yshift=10.5cm}}
\begin{axis}[
     axis on top=false,
       xmin=12, xmax=110,
       ymin=0, ymax=1.1,
       xtick={22,42,61,82,100},
      xticklabels={HD,Arr,Hep,IndLiv,Fram},
        ycomb,
        ylabel near ticks, yticklabel pos=left,
      ylabel={\textit{$R$}},
      legend style={at={(0.5,-0.15)},
      anchor=north,legend columns=3},
      ymajorgrids=true,
      grid style=dashed,
          ]
\addplot+[mark options={fill=blue},fill=blue!40!,draw=blue,  thick]
coordinates
{(17,.4516) (36,.3913) (55,.8125) (74,.7627) (93,.8837) }
\closedcycle;%
\addlegendentry{Clean Data}

\addplot+[mark options={fill=green},fill=pink,draw=green,  thick] 
coordinates
 {(20,.4193) (39,.3913) (58,.7500) (77,.7627) (96,.8801) }
\closedcycle;%
\addlegendentry{PPMD}

\addplot+[mark options={fill=white},fill=red!60!,draw=red!70!,  thick] 
coordinates
{(23,.3870) (42,.3738) (61,.7500) (80,.7179) (99,.8632) }
\closedcycle;%
\addlegendentry{PMLM \cite{R14}}
\addplot+[mark options={fill=Mustard},fill=Mustard,draw=Mustard,  thick] 
coordinates
 {(26,.3870) (45,.3846) (64,.7500) (83,.7457) (102,.8679) }
\closedcycle;%
\addlegendentry{NbAFL \cite{R29}}
\addplot+[mark options={fill=teal},fill=teal,draw=teal,  thick] 
coordinates
 {(29,.3870) (48,.3695) (67,.6666) (86,.7371) (105,.8632)}
\closedcycle;%
\addlegendentry{MLPAM \cite{R19}}
\end{axis}
\end{tikzpicture}
            \caption{ANN}
        \end{subfigure}
\caption{Recall of \textit{CM} in PPMD}
    \end{figure}

\begin{figure}[!htbp]
\centering
\begin{subfigure}[t]{0.49\textwidth}
\begin{tikzpicture}[node distance = 1cm,auto,scale=.70, transform shape]
\pgfplotsset{every axis y label/.append style={rotate=180,yshift=10.5cm}}
\begin{axis}[
      axis on top=false,
      xmin=12, xmax=110,
      ymin=0, ymax=1.1,
      xtick={22,42,61,82,100},
      xticklabels={HD,Arr,Hep,IndLiv,Fram},
        ycomb,
        ylabel near ticks, yticklabel pos=left,
      ylabel={\textit{$FS$}},
      legend style={at={(0.5,-0.15)},
      anchor=north,legend columns=3},
      ymajorgrids=true,
      grid style=dashed,
          ]
\addplot+[mark options={fill=blue},fill=blue!40!,draw=blue,  thick]
coordinates
{(17,.7404) (36,.5993) (55,.8461) (74,.8653) (93,.7895) }
\closedcycle;%
\addlegendentry{Clean Data}

\addplot+[mark options={fill=green},fill=pink,draw=green,  thick] 
coordinates
 {(20,.7199) (39,.5707) (58,.8148) (77,.8543) (96,.7752) }
\closedcycle;%
\addlegendentry{PPMD}

\addplot+[mark options={fill=white},fill=red!60!,draw=red!70!,  thick] 
coordinates
{(23,.6666) (42,.5148) (61,.7500) (80,.8431) (99,.7631) }
\closedcycle;%
\addlegendentry{PMLM \cite{R14}}
\addplot+[mark options={fill=Mustard},fill=Mustard,draw=Mustard,  thick] 
coordinates
{(26,.6756) (45,.5193) (64,.7692) (83,.8316) (102,.7697) }
\closedcycle;%
\addlegendentry{NbAFL \cite{R29}}
\addplot+[mark options={fill=teal},fill=teal,draw=teal,  thick] 
coordinates
 {(29,.6320) (48,.5134) (67,.7200) (86,.8200) (105,.7515)}
\closedcycle;%
\addlegendentry{MLPAM \cite{R19}}
\end{axis}
\end{tikzpicture}
            \caption{SVM}
        \end{subfigure}
            \hfill
\begin{subfigure}[t]{0.49\textwidth}
\begin{tikzpicture}[node distance = 1cm,auto,scale=.70, transform shape]
\pgfplotsset{every axis y label/.append style={rotate=180,yshift=10.5cm}}
\begin{axis}[
     axis on top=false,
      xmin=12, xmax=110,
      ymin=0, ymax=1.1,
      xtick={22,42,61,82,100},
      xticklabels={HD,Arr,Hep,IndLiv,Fram},
        ycomb,
        ylabel near ticks, yticklabel pos=left,
      ylabel={\textit{$FS$}},
      legend style={at={(0.5,-0.15)},
      anchor=north,legend columns=3},
      ymajorgrids=true,
      grid style=dashed,
          ]
\addplot+[mark options={fill=blue},fill=blue!40!,draw=blue,  thick]
coordinates
{(17,.7407) (36,.6620) (55,.8000) (74,.8131) (93,.7793) }
\closedcycle;%
\addlegendentry{Clean Data}
\addplot+[mark options={fill=green},fill=pink,draw=green,  thick] 
coordinates
 {(20,.7142) (39,.6475) (58,.7692) (77,.8089) (96,.7656) }
\closedcycle;%
\addlegendentry{PPMD}
\addplot+[mark options={fill=white},fill=red!60!,draw=red!70!,  thick] 
coordinates
 {(23,.6923) (42,.5477) (61,.7200) (80,.7954) (99,.7653) }
\closedcycle;%
\addlegendentry{PMLM \cite{R14}}
\addplot+[mark options={fill=Mustard},fill=Mustard,draw=Mustard,  thick] 
coordinates
 {(26,.6995) (45,.6074) (64,.7500) (83,.7954) (102,.7652) }
\closedcycle;%
\addlegendentry{NbAFL \cite{R29}}
\addplot+[mark options={fill=teal},fill=teal,draw=teal,  thick] 
coordinates
 {(29,.6873) (48,.5419) (67,.7199) (86,.7912) (105,.7558)}
\closedcycle;%
\addlegendentry{MLPAM \cite{R19}}
\end{axis}
\end{tikzpicture}
            \caption{KNN}
        \end{subfigure}
            \hfill
\begin{subfigure}[t]{0.49\textwidth}
\begin{tikzpicture}[node distance = 1cm,auto,scale=.70, transform shape]
\pgfplotsset{every axis y label/.append style={rotate=180,yshift=10.5cm}}
\begin{axis}[
      axis on top=false,
      xmin=12, xmax=110,
      ymin=0, ymax=1.1,
      xtick={22,42,61,82,100},
      xticklabels={HD,Arr,Hep,IndLiv,Fram},
        ycomb,
        ylabel near ticks, yticklabel pos=left,
      ylabel={\textit{$FS$}},
      legend style={at={(0.5,-0.15)},
      anchor=north,legend columns=3},
      ymajorgrids=true,
      grid style=dashed,
          ]
\addplot+[mark options={fill=blue},fill=blue!40!,draw=blue,  thick]
coordinates
{(17,.7096) (36,.6447) (55,.9090) (74,.7500) (93,.7796) }
\closedcycle;%
\addlegendentry{Clean Data}
\addplot+[mark options={fill=green},fill=pink,draw=green,  thick] 
coordinates
 {(20,.5925) (39,.6261) (58,.8799) (77,.7404) (96,.7761) }
\closedcycle;%
\addlegendentry{PPMD}
\addplot+[mark options={fill=white},fill=red!60!,draw=red!70!,  thick] 
coordinates
{(23,.5400) (42,.5629) (61,.8000) (80,.7250) (99,.7726) }
\closedcycle;%
\addlegendentry{PMLM \cite{R14}}
\addplot+[mark options={fill=Mustard},fill=Mustard,draw=Mustard,  thick] 
coordinates
{(26,.5562) (45,.5769) (64,.8461) (83,.7142) (102,.7723) }
\closedcycle;%
\addlegendentry{NbAFL \cite{R29}}
\addplot+[mark options={fill=teal},fill=teal,draw=teal,  thick] 
coordinates
 {(29,.5161) (48,.5470) (67,.7826) (86,.6923) (105,.7629)}
\closedcycle;%
\addlegendentry{MLPAM \cite{R19}}
\end{axis}
\end{tikzpicture}
            \caption{RF}
        \end{subfigure}
            \hfill
\begin{subfigure}[t]{0.49\textwidth}
\begin{tikzpicture}[node distance = 1cm,auto,scale=.70, transform shape]
\pgfplotsset{every axis y label/.append style={rotate=180,yshift=10.5cm}}
\begin{axis}[
      axis on top=false,
      xmin=12, xmax=110,
      ymin=0, ymax=1.1,
      xtick={22,42,61,82,100},
      xticklabels={HD,Arr,Hep,IndLiv,Fram},
        ycomb,
        ylabel near ticks, yticklabel pos=left,
      ylabel={\textit{$FS$}},
      legend style={at={(0.5,-0.15)},
      anchor=north,legend columns=3},
      ymajorgrids=true,
      grid style=dashed,
          ]
\addplot+[mark options={fill=blue},fill=blue!40!,draw=blue,  thick]
coordinates
{(17,.7333) (36,.6414) (55,.8571) (74,.6557) (93,.8335) }
\closedcycle;%
\addlegendentry{Clean Data}
\addplot+[mark options={fill=green},fill=pink,draw=green,  thick] 
coordinates
 {(20,.6896) (39,.6020) (58,.8181) (77,.6451) (96,.8220) }
\closedcycle;%
\addlegendentry{PPMD}
\addplot+[mark options={fill=white},fill=red!60!,draw=red!70!,  thick] 
coordinates
 {(23,.6666) (42,.5713) (61,.8000) (80,.5964) (99,.8071) }
\closedcycle;%
\addlegendentry{PMLM \cite{R14}}
\addplot+[mark options={fill=Mustard},fill=Mustard,draw=Mustard,  thick] 
coordinates
 {(26,.6787) (45,.5991) (64,.8148) (83,.6153) (102,.8197) }
\closedcycle;%
\addlegendentry{NbAFL \cite{R29}}
\addplot+[mark options={fill=teal},fill=teal,draw=teal,  thick] 
coordinates
 {(29,.6393) (48,.5629) (67,.7638) (86,.5937) (105,.8064)}
\closedcycle;%
\addlegendentry{MLPAM \cite{R19}}
\end{axis}
\end{tikzpicture}
            \caption{NB}
        \end{subfigure}
            \hfill
\begin{subfigure}[t]{0.49\textwidth}
\begin{tikzpicture}[node distance = 1cm,auto,scale=.70, transform shape]
\pgfplotsset{every axis y label/.append style={rotate=180,yshift=10.5cm}}
\begin{axis}[
      axis on top=false,
      xmin=12, xmax=110,
      ymin=0, ymax=1.1,
      xtick={22,42,61,82,100},
      xticklabels={HD,Arr,Hep,IndLiv,Fram},
        ycomb,
        ylabel near ticks, yticklabel pos=left,
      ylabel={\textit{$FS$}},
      legend style={at={(0.5,-0.15)},
      anchor=north,legend columns=3},
      ymajorgrids=true,
      grid style=dashed,
          ]
\addplot+[mark options={fill=blue},fill=blue!40!,draw=blue,  thick]
coordinates
{(17,.4516) (36,.4186) (55,.7647) (74,.7563) (93,.8001) }
\closedcycle;%
\addlegendentry{Clean Data}
\addplot+[mark options={fill=green},fill=pink,draw=green,  thick] 
coordinates
 {(20,.4262) (39,.4137) (58,.6923) (77,.7500) (96,.7998) }
\closedcycle;%
\addlegendentry{PPMD}
\addplot+[mark options={fill=white},fill=red!60!,draw=red!70!,  thick] 
coordinates
{(23,.4067) (42,.3913) (61,.6666) (80,.7118) (99,.7987) }
\closedcycle;%
\addlegendentry{PMLM \cite{R14}}
\addplot+[mark options={fill=Mustard},fill=Mustard,draw=Mustard,  thick] 
coordinates
 {(26,.4242) (45,.4081) (64,.6756) (83,.7333) (102,.7908) }
\closedcycle;%
\addlegendentry{NbAFL \cite{R29}}
\addplot+[mark options={fill=teal},fill=teal,draw=teal,  thick] 
coordinates
 {(29,.4067) (48,.3731) (67,.6714) (86,.6976) (105,.7908)}
\closedcycle;%
\addlegendentry{MLPAM \cite{R19}}
\end{axis}
\end{tikzpicture}
            \caption{ANN}
        \end{subfigure}
\caption{F1-score of \textit{CM} in PPMD}
    \end{figure}
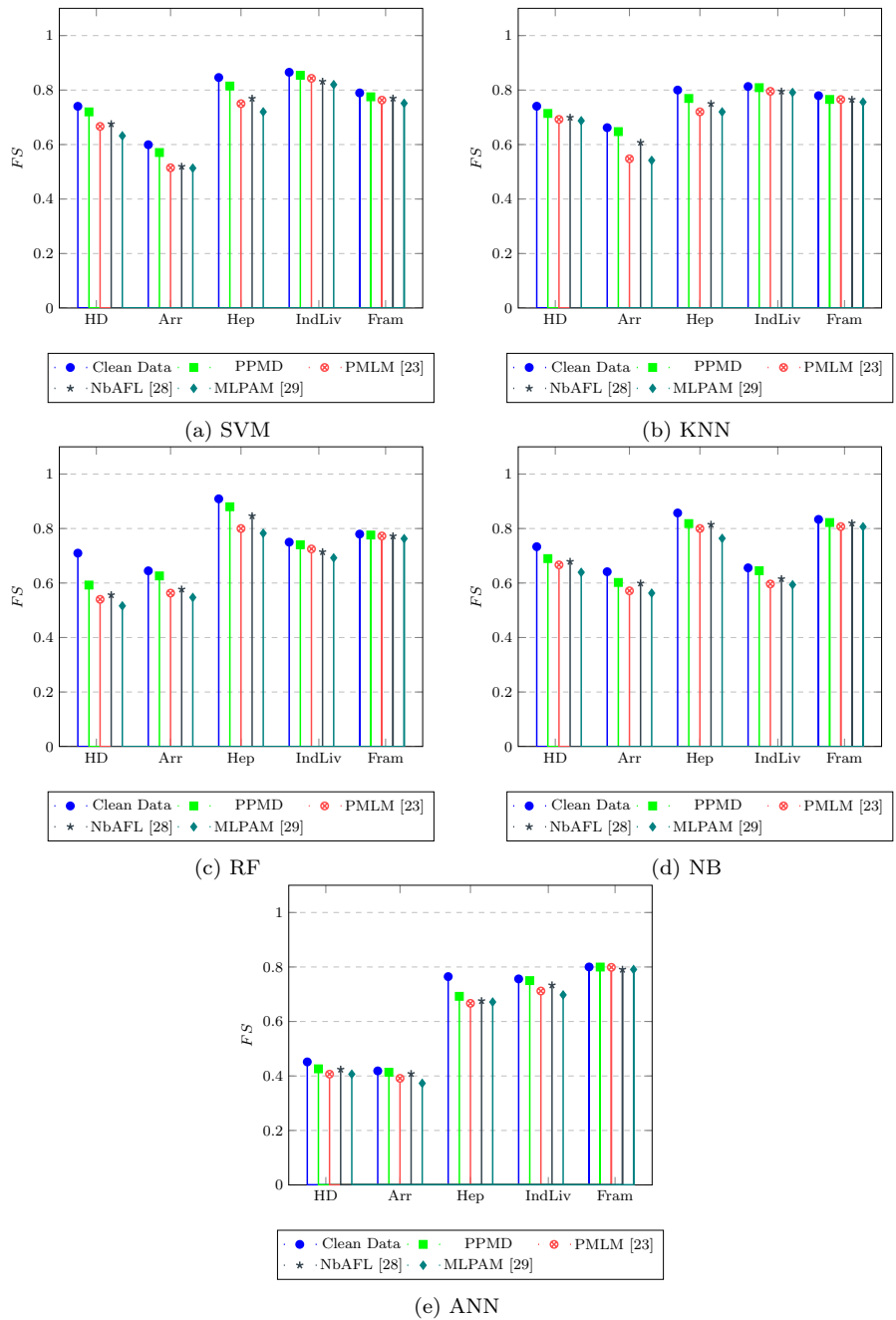
\subsection{Comparison}
The experimental results are compared with clean data, PMLM \cite{R14}, NbAFL \cite{R29} as well as MLPAM \cite{R19}, which is implemented on the same platform (Figs. 5(a)-(e) to 8(a)-(e)). The PPMD outperforms PMLM \cite{R14}, NbAFL \cite{R29}, and MLPAM \cite{R19} in all the cases because the proposed model reduces the impact of injecting noise by separating data into sensitive and non-sensitive parts. From Table 8, it is observed that the highest difference for \textit{$CA$} among PPMD, PMLM, NbAFL, and MLPAM is 15.83\% on the Hepatitis dataset using ANN classifier, and the lowest difference is found 0.0\% on the Heart Disease dataset using SVM classifier, and the Hepatits dataset using KNN classifier. 
\begin{table}[!htbp]
\caption{Improvement in the values of \textit{$CA$}, \textit{$P$}, \textit{$R$}, and \textit{$FS$} of PPMD in comparison with the values on PMLM \cite{R14}, NbAFL \cite{R29}, MLPAM \cite{R19}}
\label{tab:187}     
\begin{tabular}{p{.7cm}p{.5cm}p{.4cm}p{.4cm}p{.4cm}p{.4cm}p{.4cm}p{.4cm}p{.4cm}p{.4cm}p{.4cm}p{.4cm}p{.4cm}p{.4cm}}
\hline
Data  & Class  &\multicolumn{12}{ c }{ Decrements in the value of parameters}  \\ \cline{3-14}
set & ifier & \multicolumn{3}{ c }{\textit{$CA$}} & \multicolumn{3}{ c }{\textit{$P$}} & \multicolumn{3}{ c }{\textit{$R$}} & \multicolumn{3}{ c }{\textit{$FS$}} \\
\cline{3-14}
 &  & \cite{R14} & \cite{R29} & \cite{R19} &  \cite{R14} & \cite{R29} & \cite{R19} &  \cite{R14} & \cite{R29} & \cite{R19} &  \cite{R14} & \cite{R29} & \cite{R19}  \\ \hline
 & SVM  & 3.22 & 0.0 & 6.45 &  5.00 & 4.04 & 7.26 &  10.9 & 4.72  & 11.19  & 5.33 & 4.43 & 8.79 \\ 
& KNN & 3.23 &  3.23 & 3.23 & 1.92 & 2.17 & 5.81  & 4.16 & 2.15 & 5.37  & 2.19 & 1.47 & 2.69\\
Heart  & RF & 1.01 & 1.01 & 1.01 & 24.45 & 22.94 & 28.39 & 11.33 & 1.5 & 1.72 & 5.25 &3.63 & 7.64\\
Disease & NB  & 3.22 & 3.22 & 3.75  & 2.67 & 1.13 & 4.58  & 3.51 &2.15 & 4.37 & 2.30 & 1.09 & 5.03 \\
 & ANN  & 0.61 & 3.06 & 3.80 & 0.48 & 0.48 & 0.48  & 3.23 &3.23 & 3.23 & 1.95 &0.20 & 1.95\\ \hline
 & SVM  & 2.18 & 2.18 & 4.35 & 5.86 & 5.58 & 5.93 & 2.18 & 2.18 & 4.35 & 5.59 & 5.14 & 5.73\\ 
 & KNN  & 6.52 & 2.17 & 6.52 & 9.30 & 8.45 & 10.77  & 6.52 & 2.17 & 6.52  & 9.98 & 4.01 & 10.56\\
Arrhy & RF  & 4.35 & 2.17 & 8.70 & 3.94 & 0.93 & 5.00  & 4.35 & 4.35 & 8.70 & 6.32 & 4.92 & 7.91 \\
thmia & NB  & 1.56 & 1.56 & 3.74  & 1.13 & 1.26 & 1.76 & 1.56 & 1.56 & 3.74 & 3.07 &0.29 & 3.91 \\
 & ANN  & 0.25 & 0.25 & 1.08 & 0.0 & 0.53 & 1.05 & 1.75 & 0.67 & 2.18 & 2.24 & 0.56 & 4.06\\ \hline
 & SVM & 6.25 & 6.25 & 6.25 & 4.10 & 4.58 & 4.58 & 9.85 & 4.16 & 10.41 & 6.48 & 4.56 & 9.48\\ 
 & KNN & 6.25 & 0.0 & 0.0 & 4.76 & 2.19 & 2.19 & 1.52 &1.52 & 2.08 & 4.92 & 1.92 & 4.93\\
Hepa & RF & 12.50 & 6.25 & 12.50 & 1.65 & 1.65 & 11.91 & 1.28 & 3.36 & 9.61 & 7.99 & 3.38 & 9.73\\
titis & NB & 6.25 & 6.25 & 6.25 & 6.41 & 4.76 & 11.97 & 5.50 &1.92 & 1.92 & 1.81 & 0.33 & 5.43\\
 & ANN & 15.11 & 8.0 & 15.83 & 4.28 & 2.48 & 5.95 & 0.0 & 0.0 & 8.34 & 2.57 & 1.67 & 2.09\\ \hline
 & SVM & 1.69 & 3.39 & 5.08 & 1.69 & 3.39 & 6.15 & 0.0 & 0.0 & 2.44 & 1.12 & 2.27 & 3.43\\ 
Indian & KNN & 1.69 & 1.69 & 3.39 & 1.67 & 2.75 & 3.26 & 0.0 & 0.39 & 0.75 & 1.35 & 1.35 & 1.77\\
Liver & RF & 1.69 & 3.39 & 5.08 & 6.01 & 0.24 & 4.60 & 0.0 & 1.06 & 4.3 & 1.54 & 2.62 & 4.81\\
Patient  & RF & 1.70 & 1.70 & 3.39 & 2.45 & 1.81 & 5.23 & 2.77 & 0.05 & 7.03 & 4.87 & 2.98 & 5.14\\
  & ANN & 3.05 & 1.46 & 3.24 & 9.49 & 3.77 & 9.49 & 4.48 & 1.70 & 2.56 & 3.82 & 1.67 & 5.24\\ \hline
 & SVM & 0.71 & 0.24 & 2.12 & 2.76 & 1.18 & 4.09 & 0.71 & 0.24 & 2.12 & 1.21 & 0.55 & 2.37\\ 
 & KNN & 0.71 & 0.24 & 1.18 & 0.90 & 0.74 & 2.29 & 0.71 & 0.24 & 1.18  & 0.03 & 0.04 & 0.98\\
Frami & RF & 1.18 & 0.71 & 2.13 & 2.04 & 1.77 & 3.24 & 1.18 & 0.71 & 2.13 & 0.35 & 0.38 & 1.32\\
ngham & NB & 0.95 & 0.24 & 2.13 & 1.62 & 0.28 & 2.26 & 0.95 & 0.24 & 2.13 & 1.49 & 0.23 & 1.56\\
 & ANN & 0.24 & 1.13 & 9.23 & 0.09 & 1.08 & 1.29 & 1.69 & 1.22 & 1.69 & 0.11 & 0.90 & 0.90\\
\hline
\end{tabular}
\end{table}
Likewise, the maximum gap for \textit{$P$} is 28.39\% on the Heart Disease dataset using the RF classifier, but the lowest difference is found 0.0\% on the Arrhythmia dataset using the ANN classifier. The \textit{$R$} of PPMD maximum improved by 11.33\% from PMLM, NbAFL, and MLPAM on the Heart Disease dataset using the RF classifier, whereas the smallest improvement is 0.0\%  on the Hepatitis dataset using ANN classifier, and on the Indian-liver-patient dataset using the SVM, KNN, RF classifiers. The highest difference for \textit{$FS$} is 10.56\% on the Arrhythmia dataset using the KNN classifier, while the lowest difference for \textit{$FS$} is 0.03\% on the Framingham dataset using the KNN classifier.
\par
Moreover, the results of PPMD are less than the results of Clean data in all the cases due to noise addition. Table 9 shows that the maximum gap for \textit{$CA$} between PPMD and clean data is 6.25\% on the Hepatitis dataset using the SVM, KNN, RF, and NB classifiers, but the smallest gap is found 0.12\% on the Framingham dataset using the ANN classifier. Similarly, the highest difference for \textit{$P$} is 8.09\% on the Hepatitis dataset using the RF classifier, while the lowest difference is found 0.0\% on the Hepatitis dataset using the KNN classifier. The \textit{$R$} of PPMD the maximum decrement by 8.72\% from clean data on the Hepatitis dataset using the RF classifier. The smallest decrement is 0.0\% on the Arrhythmia dataset using the ANN, Hepatitis dataset using SVM, and Indian-liver-patient dataset using SVM and ANN classifiers. The highest difference for \textit{$FS$} is 11.71\% on the Heart Disease dataset using the RF classifier, but the lowest difference for \textit{$FS$} is 0.03\% on the Framingham dataset using the ANN classifier. But still, the results of PPMD are almost equal and also offer more protection compared to the clean data.
\begin{table}[!hbp]
\caption{Reduction in the values of \textit{$CA$}, \textit{$P$}, \textit{$R$}, and \textit{$FS$} of PPMD in comparison to the values on clean data  }
\label{tab:234}      
\begin{tabular}{llllll}
\hline\noalign{\smallskip}
Dataset & Classifier & \multicolumn{4}{ c }{\% decrement in the value of parameters}  \\ \cline{3-6}
& & \textit{$CA$} & \textit{$P$} & \textit{$R$} & \textit{$FS$} \\
\noalign{\smallskip}\hline\noalign{\smallskip}
 & SVM & 3.23 & 1.92 & 2.19 & 2.05 \\ 
 & KNN & 3.22 & 3.08 & 4.76 & 2.65 \\
Heart Disease & RF & 2.21 & 4.61 & 7.78 & 11.71 \\
 & NB & 3.23 & 3.58 & 3.92 & 4.37 \\
 & ANN & 0.61 & 1.83 & 3.23 & 2.54 \\ \hline
 & SVM & 2.17 & 3.07 & 2.17 & 2.86 \\ 
 & KNN & 2.18 & 2.28 & 2.18 & 1.45 \\
Arrhythmia & RF & 2.17 & 3.80 & 2.17 & 1.86 \\
 & NB & 1.75 & 3.84 & 1.75 & 3.94 \\
 & ANN & 0.65 & 1.10 & 0.00 & 0.49 \\ \hline
 & SVM & 6.25 & 5.24 & 0.00 & 3.13 \\ 
 & KNN & 6.25 & 0.00 & 7.57 & 3.08 \\
Hepatitis & RF & 6.25 & 8.09 & 8.72 & 2.91 \\
 & NB & 6.25 & 6.67 & 4.89 & 3.90 \\
 & ANN & 2.63 & 7.94 & 6.25 & 7.24 \\ \hline
 & SVM & 1.70 & 1.70 & 0.00 & 1.10 \\ 
 & KNN & 1.70 & 2.99 & 0.37 & 0.42 \\
Indian Liver & RF & 1.70 & 0.97 & 1.67 & 0.96 \\
 Patient & NB & 1.69 & 0.33 & 1.28 & 1.06 \\
 & ANN & 3.05 & 1.23 & 0.00 & 0.63 \\ \hline
 & SVM & 0.71 & 0.37 & 0.71 & 1.43 \\ 
 & KNN & 0.71 & 0.95 & 0.71 & 1.37 \\
Framingham & RF & 0.41 & 0.21 & 0.41 & 0.35 \\
 & NB & 1.41 & 1.17 & 1.41 & 1.15 \\
 & ANN & 0.12 & 0.02 & 0.05 & 0.03 \\ 
\noalign{\smallskip}\hline
\end{tabular}
\end{table}
\subsection{Statistical analysis}
Statistical analysis is used to validate the \textit{$CA$}, \textit{$P$}, \textit{$R$}, and \textit{$FS$} of the proposed model. In this context, the non-parametric test is applied to the dataset that is not normally distributed. The null hypothesis states that the acquired results from different methods are statistically identical in the Wilcoxon signed-rank test. This test compares the performance of PPMD model to that of the existing PMLM \cite{R14}, NbAFL \cite{R29}, and MLPAM \cite{R19} models. The test is run on the dataset with a significance level \cite{R28} (p-value) of 0.05 to determine the importance of classifying parameters. Table 10 demonstrates the results of the test statistics.
 \par
While comparing PPMD and PMLM \cite{R14}, it is observed that the null hypothesis for \textit{$CA$}, \textit{$P$}, \textit{$R$}, and \textit{$FS$} is rejected because their p-values are less than 0.05, indicating that PPMD for the Heart Disease dataset is valid. The null hypothesis is rejected for \textit{$CA$}, \textit{$R$}, and \textit{$FS$} but accepted for \textit{$P$} on the Arrhythmia dataset. The null hypothesis is rejected for \textit{$CA$}, \textit{$P$}, and \textit{$FS$}, whereas accepted for \textit{$R$} on the Hepatitis, and Indian-Liver-Patient datasets. The null hypothesis is rejected for  \textit{$CA$}, \textit{$P$}, \textit{$R$}, and \textit{$FS$} on the Framingham dataset. 
\par
Similarly, comparing PPMD, and  NbAFL \cite{R29}, the null hypothesis is rejected for \textit{$P$}, \textit{$R$}, and \textit{$FS$} but accepted for \textit{$CA$} on the Heart Disease dataset. The null hypothesis is rejected for \textit{$CA$}, \textit{$P$}, \textit{$R$}, and \textit{$FS$} on the Arrhythmia dataset. The null hypothesis is rejected for \textit{$P$}, and \textit{$FS$} but accepted for \textit{$CA$}, and \textit{$R$} on the Hepatitis dataset. The null hypothesis is rejected for \textit{$CA$}, \textit{$P$}, and \textit{$FS$}, whereas accepted for \textit{$R$} on Indian-live-Patient dataset. The null hypothesis is rejected for \textit{$CA$}, \textit{$P$}, \textit{$R$}, and \textit{$FS$} on the Framingham dataset. 
\par
Moreover, comparing PPMD and  MLPAM \cite{R19}, the null hypothesis is rejected for \textit{$CA$}, \textit{$P$}, \textit{$R$}, and \textit{$FS$} on the Heart Disease and Arrhythmia datasets. The null hypothesis is rejected for \textit{$P$}, \textit{$R$}, and \textit{$FS$} but accepted for \textit{$CA$} on the Hepatitis dataset. The null hypothesis is rejected for \textit{$CA$}, \textit{$P$}, \textit{$R$}, and \textit{$FS$} on the Indian-Liver-Patient and Framingham datasets. 
\begin{table}[!htbp] 
\centering
\caption{Wilcoxon test statistics (p-value is 0.05)}
\label{table:267}
\begin{tabular}{p{.7cm}p{1.4cm}llllll}
\hline
Data & Classification & \multicolumn{2}{ c }{Comparison of} & \multicolumn{2}{ c }{Comparison of} & \multicolumn{2}{ c }{Comparison of} \\  
set & Parameters & \multicolumn{2}{ c }{PPMD, PMLM \cite{R14}} & \multicolumn{2}{ c }{PPMD, NbAFL \cite{R29}} &  \multicolumn{2}{ c }{PPMD, MLPAM \cite{R19}}  \\    \cline{3-8}
& & p-value & Result & p-value & Result & p-value & Result\\ \hline 
& \textit{$CA$}  & 0.042 & RE & 0.68 & AC & 0.43 & RE \\ 
Heart & \textit{$P$} & 0.043 & RE & 0.43 & RE & 0.43 & RE\\ 
Disease & \textit{$R$} & 0.043 & RE & 0.42 & RE & 0.43 & RE\\ 
& \textit{$FS$} & 0.043 & RE & 0.43 & RE & 0.43 & RE\\ \hline
& \textit{$CA$}  & 0.043 & RE & 0.42 & RE & 0.43 & RE\\ 
Arrhy & \textit{$P$}  & 0.068 & AC  & 0.43 & RE & 0.43 & RE\\ 
thmia & \textit{$R$} & 0.043 & RE & 0.43 & RE & 0.43 & RE\\ 
& \textit{$FS$} & 0.043 & RE  & 0.43 & RE & 0.43 & RE\\ \hline
& \textit{$CA$}  & 0.039 & RE & 0.59 & AC & 0.66 & AC\\ 
Hepa & \textit{$P$}  & 0.043 & RE & 0.43 & RE & 0.43 & RE\\ 
titis & \textit{$R$} & 0.068 & AC & 0.68 & AC & 0.43 & RE\\ 
& \textit{$FS$}  & 0.043 & RE & 0.43 & RE & 0.43 & RE\\ \hline
& \textit{$CA$} & 0.042 & RE & 0.43 & RE & 0.41 &  RE\\ 
Indian & \textit{$P$} & 0.043 & RE & 0.43 & RE & 0.43 & RE\\ 
Liver & \textit{$R$} & 0.180 & AC & 0.68 & AC & 0.43 & RE\\ 
Patient & \textit{$FS$} & 0.043 & RE & 0.43 & RE & 0.43 & RE\\ \hline
& \textit{$CA$} & 0.043 & RE & 0.42 & RE & 0.42 & RE\\ 
Frami & \textit{$P$} & 0.043 & RE & 0.43 & RE & 0.43 & RE\\ 
ngham & \textit{$R$} & 0.043 &  RE & 0.42 & RE & 0.42 & RE\\ 
 & \textit{$FS$} & 0.043 & RE & 0.43 & RE & 0.43 & RE\\ \hline
 \noalign{\smallskip}
\end{tabular}
AC: The null hypothesis is accepted, RE: The null hypothesis is rejected
\end{table}
\section{Conclusion and future work}
This paper proposed a novel model named PPMD that preserves the privacy of outsourced sensitive data provided by various data owners in a real cloud environment. PPMD allows multiple data owners to outsource their data to the cloud for storing and computation. In this work, data owners added different statistical noise to sensitive data according to their queries for data protection. The cloud service provider has also provided the classification service. The experiments have been conducted, and results show that PPMD ensures high accuracy, precision, recall, and F1-score improvement up to 29\% over the existing works. The model's performance over the well-known data sets and comparison with existing works showed that PPMD is more secure, efficient, and optimal. The future aim of this work would be to share collected data among requesting users and devise a more efficient privacy-preserving mechanism to protect the data for various owners. 
\begin{acknowledgements}
This work is supported by University Grant Commission, New Delhi, India under the scheme of National Eligibility Test-Junior Research Fellowship (NET-JRF) with reference id-3515/(NET-NOV 2017).
\end{acknowledgements}

 \section*{Compliance with ethical standards}
\textbf{Conflict of interest} The authors have no conflict of interest regarding the publication. \\ \\
\textbf{Ethical approval} This article does not contain any studies with human participants or animals performed by any of the authors. \\ \\
\textbf{Informed consent} Informed consent was obtained from all individual participants included in the study.\\ \\
\textbf{Authors’ contribution} Both the authors have discussed and constructed the ideas, designed the privacy-preserving model, and wrote the paper together.

\end{document}